\documentclass[prb,showpacs,floatfix,preprint,amsmath,amssymb]{revtex4}

\usepackage{graphicx}
\usepackage{dcolumn}
\usepackage{bm}
\usepackage{epsfig}

\def\be{\begin{eqnarray}}
\def\ee{\end{eqnarray}}
\def\ba{\begin{eqnarray}}
\def\ea{\end{eqnarray}}
\def\bm{\begin{math}}
\def\me{\end{math}}

\begin{document}

\title{Spinodal Decomposition in Thin Films: Molecular Dynamics Simulations 
of a Binary Lennard-Jones Fluid Mixture}
\author{Subir K. Das$^{\text{1}}$}
\author{Sanjay Puri$^{\text{2}}$}
\author{J\"urgen Horbach$^{\text{1}}$}
\author{Kurt Binder$^{\text{1}}$}
\affiliation{$^1$Institut f\"ur Physik,
             Johannes Gutenberg--Universit\"at Mainz,
             D--55099 Mainz, GERMANY. \\
             $^2$School of Physical Sciences, 
             Jawaharlal Nehru University, 
             New Delhi--110067, INDIA.}
\date{\today}
\begin{abstract}
We use molecular dynamics (MD) to simulate an unstable homogeneous
mixture of binary fluids (AB), confined in a slit pore of width $D$.
The pore walls are assumed to be flat and structureless, and attract
one component of the mixture (A) with the same strength.  The pair-wise
interactions between the particles is modeled by the Lennard-Jones
potential, with symmetric parameters that lead to a miscibility gap in
the bulk.  In the thin-film geometry, an interesting interplay occurs
between surface enrichment and phase separation.

We study the evolution of a mixture with equal amounts of A and B,
which is rendered unstable by a temperature quench.  We find that
A-rich surface enrichment layers form quickly during the early stages
of the evolution, causing a depletion of A in the inner regions of the
film. These surface-directed concentration profiles propagate from the
walls towards the center of the film, resulting in a transient layered
structure. This layered state breaks up into a columnar state, which
is characterized by the lateral coarsening of cylindrical domains. The
qualitative features of this process resemble results from previous
studies of diffusive Ginzburg-Landau-type models [S.~K. Das, S. Puri,
J. Horbach, and K. Binder, Phys. Rev. E {\bf 72}, 061603 (2005)],
but quantitative aspects differ markedly. The relation to spinodal
decomposition in a strictly 2-$d$ geometry is also discussed.
\end{abstract}

\maketitle

\section{Introduction}
Thin fluid films have a broad range of applications in technology as
lubricants, protecting layers, for production processes of layered
structures in microelectronics, etc. In particular, ultra-thin films
have become extremely important in the context of nanotechnology
\cite{1,2,3,4,5,6,7}. The interplay of surface effects and finite-size
effects with the bulk behavior of these systems poses challenging
theoretical problems \cite{8,9,10,11}.  A particularly interesting problem
in this context is the phase separation of binary (or multi-component)
mixtures in thin films, as most materials of practical interest have
more than one component, e.g., metallic alloys, ceramics, polymer
blends, etc. Phase changes in reduced geometries (e.g., 2-$d$ systems)
differ in many aspects from the bulk behavior in three dimensions. The
interplay between surface and bulk behavior leads to complex phenomena
such as wetting transitions, prewetting and layering transitions,
etc. \cite{12,13,14,15,16,17,18,19,20}. These transitions may compete
with phase changes that occur in the bulk, such as phase separation
in mixtures \cite{21,22,23,24,25,26}. There has been intense study of
phenomena such as {\it surface-directed spinodal decomposition} (SDSD) or
{\it surface-directed phase separation} \cite{27,28,29,30,31,32,33,34,35},
but our theoretical understanding of these problems is still incomplete.

\begin{figure} 
\centering
\includegraphics*[width=0.9\textwidth]{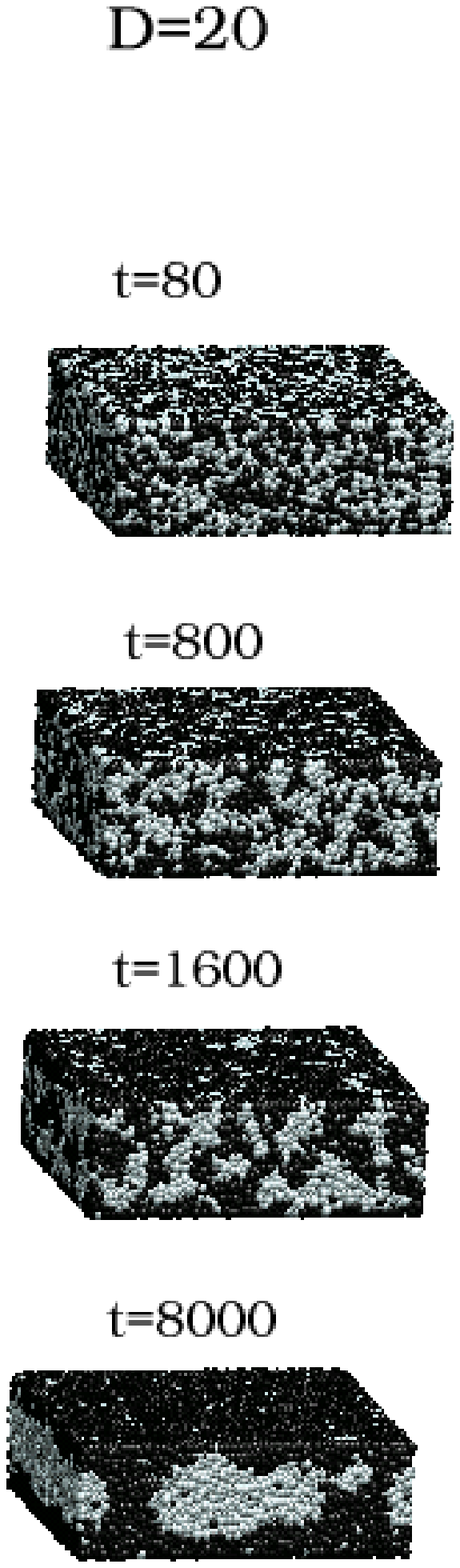}
\vspace*{-3.5cm}
\caption{\label{fig1}
Snapshot pictures of surface-directed spinodal decomposition (SDSD)
in a binary (AB) Lennard-Jones (LJ) mixture, which is confined in an
$L \times L \times D$ thin-film geometry, with $L=64$, $D=20$. (All lengths
are measured in units of the LJ diameter.) Periodic boundary conditions
were applied in the lateral directions, while the impenetrable $L
\times L$ surfaces (representing the walls of a slit pore) attract the
A-particles. The initial condition for this run consisted of a random
mixture of equal amounts of A and B, corresponding to a critical quench.
Time is also measured in dimensionless LJ units -- see
Sec. III, where further details of the simulation are specified. The
A-particles are marked black, and the B-particles are marked gray.
The system quickly develops concentration inhomogeneities ($t=800$), in
particular A-rich layers form rapidly at the walls ($t=800$, $t=1600$).
The late stages of phase separation are characterized by the lateral
coarsening of columnar structures ($t=8000$).}
\end{figure}

To gain a better understanding of SDSD in thin fluid films, we have
undertaken a comprehensive molecular dynamics (MD) simulation of
binary mixtures in a slit pore. A preliminary account of our results
has been published as a letter \cite{dphb}. In this paper, we present
detailed results from this study.  Our MD simulations are based on a
symmetric binary (AB) Lennard-Jones (LJ) mixture, for which the bulk
phase diagram has been determined to high accuracy \cite{36}. We also
have a good understanding of various other properties of this mixture,
e.g., static and dynamic response and correlation functions, transport
coefficients \cite{36,37}, and the interfacial tension between coexisting
A-rich and B-rich phases \cite{38}. Many previous simulation studies
of phase-separation kinetics in a thin-film geometry \cite{39,40} have
used Ginzburg-Landau (GL) models, and hence lack any direct connection
to a microscopic description. Our present modeling bridges the gap
between the atomistic description of liquids and the mesoscale domain
structures that form as the kinetics of phase separation proceeds
(see Fig.~\ref{fig1}). This direct approach has the further merit
that hydrodynamic interactions are automatically incorporated.  It is
well-known that these interactions have a pronounced effect on the
kinetics of domain growth \cite{21,22,23,24,25,26,41,42,43,44,45}.

In order to account for hydrodynamics in the framework of GL models,
rather extensive computations are required \cite{46,47,48,49,50}. The
GL approach is appropriate if one is primarily concerned
with the scaling behavior of the late-stage domain growth
\cite{21,22,23,24,25,26}. But recent work based on lattice
Boltzmann simulations questions the quantitative validity of GL models,
and reports very slow crossovers extending over many decades in time
\cite{53} (however, a ``somewhat narrower'' crossover region was reported in a
subsequent work \cite{53a}). This study also raises questions about some of the previous
work on this problem, using the lattice Boltzmann method or related
``lattice gas''-approximations to the Navier-Stokes equations of
hydrodynamics \cite{54,55,56,57,58}. However, the lattice Boltzmann
description is even more remote from an atomistic description of matter
than the GL model. Further, it is not clear how one incorporates the
proper boundary conditions with respect to complete or partial wetting in
such an approach \cite{59,60}. The studies mentioned above are primarily
concerned with domain growth in $d=3$. For the $d=2$ case, it is even
controversial \cite{61,62,63} to what extent a scaling behavior describes
the late stages of coarsening. In view of these problems, it is useful to
undertake an MD simulation despite the fact that the accessible scales in
length and time are limited. Earlier MD studies of bulk phase separation
\cite{64,65} have addressed coarsening in $d=3$ and in $d=2$, but the
latter work is somewhat inconclusive \cite{66}.  Further, there have
only been preliminary MD studies of SDSD in thin films \cite{mk93,st99}.

The outline of this paper is as follows. In Sec.~II, the theoretical
background (equilibrium phase behavior of binary mixtures in a thin-film
geometry, theory of domain growth, etc.) will be concisely reviewed.
In Sec.~III, we provide details of our MD methods. In Sec.~IV, we present
simulation results for SDSD in thin films.  Finally, Sec.~V concludes
this paper with a summary and discussion, including a comparison to the
GL approach.

\section{Theoretical Background}
\subsection{Equilibrium phase behavior of binary mixtures confined
between walls}
\begin{figure}[t] 
\vspace*{-1.5cm}
\centering
\includegraphics*[width=0.7\textwidth]{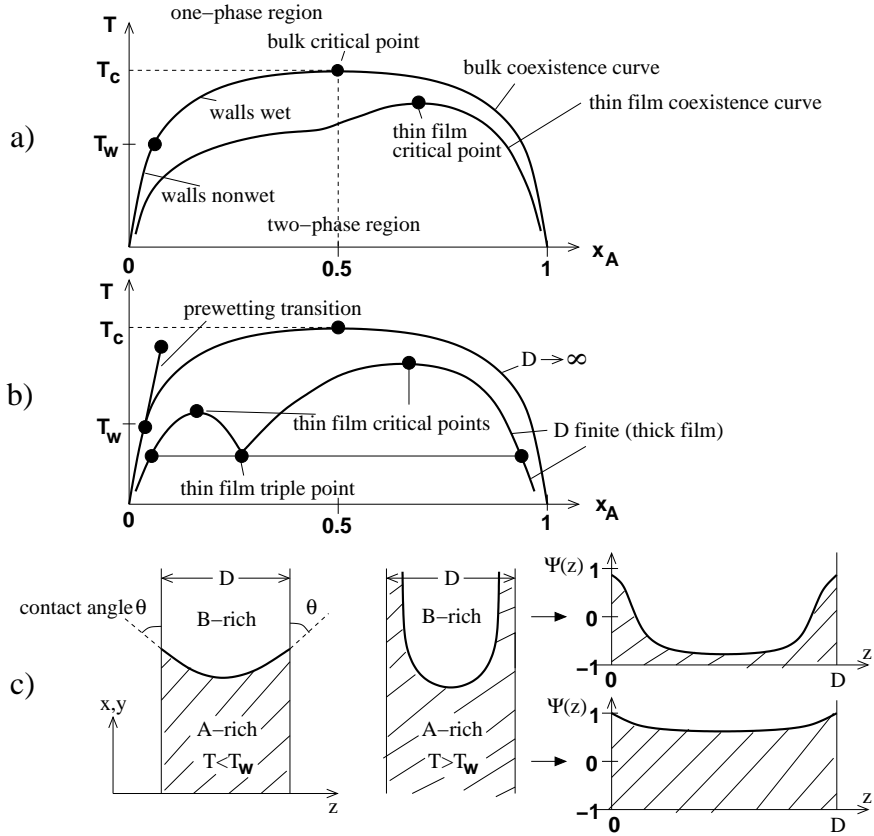}
\vspace*{-0.5cm}
\caption{\label{fig2}
Schematic phase diagrams (a,b) and corresponding states (c) of a symmetric
AB mixture in a thin film of thickness $D$. The film is symmetric,
viz., both walls attract the A-particles with the same strength. We
emphasize that the wetting transition only occurs in the limit $D
\rightarrow \infty$.  For finite $D$, the transition of the walls from
nonwet (or partially wet) to wet (or completely wet) is rounded into a
smooth gradual change. This transition is of second order in (a), while
(b) refers to a first-order wetting transition. In (b), a prewetting
transition line exists in the one-phase region, with one end being a
prewetting critical point at high temperatures. The other end of this line
is at the wetting transition temperature $T_{\rm w}$, at the coexistence
curve that separates the two-phase region from the one-phase region. Note
that the critical concentration of a symmetric binary mixture is $x^{\rm
crit}_{\rm A}=0.5$ in the bulk, but is shifted to a larger value $x_{\rm
A}$ in the thin film.  Further, the critical temperature of the film
typically is lower than in the bulk, $T_{\rm c}(D)< T_{\rm c}(\infty)
= T_{\rm c}$. For the case of first-order wetting and large enough $D$,
a thin-film analog of the prewetting transition exists, as evidenced by
the thin-film critical point at the left side of the phase diagram. When
the thin enrichment layer segregation meets the lateral segregation of the
``thick'' film, a thin-film triple point occurs at a temperature close
to $T_{\rm w}$. For thin films, this triple point and the left critical
point may merge and annihilate each other, and then the corresponding
phase diagram is similar to that in (a). In (c), we provide schematic
pictures of the thin-film states in the case of lateral phase separation.}
\end{figure}

A homogeneous binary mixture becomes unstable to phase separation when
it is quenched into the miscibility gap (see Fig.~\ref{fig2}).  For a
symmetric mixture, the miscibility gap is symmetric with respect to the
concentration $x^{\rm crit}_{\rm A}=1/2$.

At the surface of a semi-infinite mixture, one may encounter a wetting
transition \cite{12,13,14,15,16,17,18,19,20}. This transition implies a
singular behavior of the surface excess free energy $F_{\rm S}$, which
is defined as (for a film between two walls at distance $D$)
$F_{\rm film}=F_{\rm b} + 2 F_{\rm S}/D$, $D \rightarrow \infty$,
$F_{\rm b}$ being the bulk free energy of the system. Assuming, as done
in Fig.~\ref{fig2}, that the wetting transition occurs at the surface of
B-rich mixtures (caused by the preferential attraction of A-particles to
the walls), the transition is characterized by a divergence of the surface
excess concentration of A, $x^{\rm surf}_{\rm A}$. This quantity can be
obtained from $F_{\rm S}$ via suitable derivatives, or by integrating
the concentration profile \cite{12,13,14,15,16,17,18,19,20,70}
\begin{equation} \label{eq3}
x^{\rm surf}_{\rm A} = \int\limits^{D/2}_0 \left[x_{\rm A} (z) -
x^{(1)}_{\rm A, coex} \right] dz \quad , \quad D \rightarrow \infty
\quad ,
\end{equation}
where $z$ is the distance from the wall, which is located at
$z=0$. If the wall is {\it nonwet} (or {\it partially wet})
\cite{12,13,14,15,16,17,18,19,20}, $x^{\rm surf}_{\rm A}$ tends
to a finite value ($x^{\rm surf}_{\rm A, coex}$) when $x_{\rm A}
\rightarrow x^{(1) }_{\rm A, coex}$ from the one-phase region. On the
other hand, for a {\it wet} (or {\it completely wet}) wall, $x^{\rm
surf}_{\rm A} = \infty$ -- corresponding to an infinitely thick A-rich
wetting layer coating the wall, separated from the B-rich bulk by a
flat interface.

At the coexistence curve $x^{(1)}_{\rm A, coex}$, the surface excess free
energy is that of an A-rich phase $F^{\rm B-rich}_{\rm S, coex}$ if the
wall is nonwet. For a wet wall, we have $F_{\rm S}=F^{\rm A-rich}_{\rm
S, coex} + \gamma_{\rm AB}$, $\gamma_{\rm AB}$ being the interfacial
tension between coexisting A-rich and B-rich phases. These quantities
also determine the contact angle $\theta$ at which an A-B interface in
the nonwet region meets the wall \cite{12,13,14,15,16,17,18,19,20}
\begin{equation} \label{eq4}
\cos \theta= \frac{\left(F^{\rm B-rich}_{\rm S, coex} -
F^{\rm A-rich}_{\rm S, coex}\right)}{\gamma_{\rm AB}} \quad , \quad {\rm if}
\quad F^{\rm B-rich}_{\rm S, coex} < F^{\rm A-rich}_{\rm S, coex} +
\gamma_{\rm AB} \quad .
\end{equation}

If the state of the system is changed such that one increases the
temperature but stays always at the coexistence curve $x^{(1)}_{\rm A,
coex}$, one encounters a wetting transition at temperature $T_{\rm w}$
(Fig.~\ref{fig2}), where the state of the wall changes from nonwet
$(T<T_{\rm w})$ to wet $(T>T_{\rm w})$. This transition may be of second
order (Fig.~\ref{fig2}a) or first order (Fig.~\ref{fig2}b). In the
second-order case, $x^{\rm surf}_{\rm A}$ diverges continuously when
$T \rightarrow T_w^-$, while otherwise there is a discontinuous jump
in $x^{\rm surf}_{\rm A}$ from a finite value at $T_w^-$ to $\infty$ at
$T_w^+$. In the first-order case, there is also a prewetting transition
in the one-phase region (Fig.~\ref{fig2}b), where the thickness of the
A-rich surface layer jumps from a smaller value to a larger (but finite)
value. This line of prewetting transitions ends in a prewetting critical
point.

This brief review of wetting phenomena provides the basis to
understand the equilibrium behavior of binary mixtures in thin films
\cite{71,72,73,74,75,76,77,78,79,80}. If the walls are neutral (i.e.,
it has the same attractive interactions with both A-particles and
B-particles), the critical concentration remains $x^{\rm crit}_{\rm A}
=1/2$. However, the critical temperature $T_{\rm c}(D)$ is lowered
\cite{71,76,77} relative to the bulk:
\begin{equation} \label{eq5}
T_{\rm c}-T_{\rm c}(D) \propto D ^{- 1/ \nu} ,
\end{equation}
where $\nu \simeq 0.629$ \cite{68,69} is the critical exponent of
the correlation length $\xi$ of concentration fluctuations (in the
universality class of the $d=3$ Ising model). Note, however, that critical
correlations at fixed finite $D$ can become arbitrarily long-range only
in the lateral direction parallel to the film. Thus, the transition at
$T_{\rm c}(D)$ belongs to the class of the $d=2$ Ising model.  The states
below the coexistence curve of the thin film correspond to two-phase
equilibria characterized by lateral phase separation.

When there is a preferential attraction of A-particles to the walls,
the phase diagram of the thin film is no longer symmetric with respect
to $x_{\rm A}=1/2$, although we did assume such a symmetry in the
bulk. The shift of $x_{\rm A}^{\rm crit}$ and the resulting change of
the coexistence curve, is the analog of {\it capillary condensation of
gases} \cite{74,81} for binary mixtures.

The coexisting phases in the region below the coexistence curve of the
thin film are inhomogeneous in the direction perpendicular to the walls
(see Fig.~\ref{fig2}c). In the A-rich phase, we expect only a slight
enhancement of the order parameter $\psi(z)$, which is defined in terms
of the densities $n_{\rm A}(z)$, $n_{\rm B} (z)$ of A and B particles as
\begin{equation} \label{eq7}
\psi(z)=\frac{n_{\rm A} (z) - n_{\rm B} (z)}{n_{\rm A} (z) + n_{\rm
B} (z)} \quad .
\end{equation}
In the B-rich phase, however, we expect pronounced enrichment layers. As
$D \rightarrow \infty$, the thickness of these layers diverges for $T>T_{\rm w}$
but stays finite for $T<T_{\rm w}$.  In a film of finite thickness, the width
of A-rich surface layers also stays finite, e.g., for $T>T_{\rm w},~x_{\rm
A}^{\rm surf} \propto \ln D$ for short-range surface forces, while $x^{\rm
surf}_{\rm A} \propto D^{1/3}$ for non-retarded van der Waals' forces
\cite{76,82}. Thus, the wetting transition is always rounded off in a
thin film.  The prewetting line (Fig.~\ref{fig2}b) does have an analog in
films of finite thickness $D$, for sufficiently large $D$. This transition
splits into a two-phase region at small $x_{\rm A}$ between the thin-film
triple point and the thin-film critical point on the B-rich side. This
two-phase region corresponds to a coexistence between B-rich phases
with A-rich surface layers, both of which have finite (but different)
thickness. As $D \rightarrow \infty$, the thin-film critical point on
the B-rich side moves into the prewetting critical point, while the
thin-film triple point merges with the first-order wetting transition.
On the other hand, when $D$ becomes small, the thin-film critical point
and the thin-film triple point may merge and annihilate each other.
For still smaller $D$, the thin-film phase diagram then has the shape
shown in Fig.~\ref{fig2}a, although one has first-order wetting in the
semi-infinite bulk (Fig.~\ref{fig2}b).

Finally, we comment on the state encountered below the bulk coexistence
curve, but above the coexistence curve of the thin film. When one crosses
the bulk coexistence curve, there is a {\it rounded transition} towards a
layered (stratified) structure with two A-rich layers at the walls and a
B-rich layer in the middle. The temperature range over which this rounded
transition is smeared is also of order $\Delta T \propto D^{-1/\nu}$
around $T_{\rm c}$.  Hence, for large $D$, this segregation in the
direction normal to the walls may easily be mistaken (in experiments or
simulations) as a true (sharp) phase transition. We stress that this is
not a true transition -- one is still in the one-phase region of the thin
film, although the structure is strongly inhomogeneous!  The situation
qualitatively looks like the concentration profile shown in the upper part
of Fig.~\ref{fig2}c. The difference is that, for $D \rightarrow \infty$,
the thickness of true wetting layers scales sub-linearly with $D$,
as noted above. However, for phase separation in the normal direction
which gradually sets in when one crosses the bulk coexistence curve,
one simply has A-rich domains of macroscopic dimensions (proportional
to $D$) adjacent to both walls.  Unfortunately, the layers resulting in
this stratified structure are often referred to as ``wetting layers''
in the literature, although this is completely misleading. We reiterate
that A-rich wetting layers only form when a B-rich domain extends to
the surface, which is not the case here.

We also caution the reader that a picture in terms of A-rich layers
at the walls and a B-rich domain in the inside of the film is an
over-simplification because the thickness of the domain walls cannot
really be neglected in the region $T_{\rm c}(D)< T<T_{\rm c}$, where
a stratified structure occurs in equilibrium. This is seen from the
relation $\xi \propto (1-T/T_{\rm c})^{-\nu}$, in conjunction with
Eq.~(\ref{eq5}), which shows that $\xi \sim O(D)$ at $T_{\rm c}(D)$.
Thus, domains and domain walls are not well-distinguished in the region
under consideration, since the interfacial width is $O(\xi)$ \cite{20,70}.

When the interface between A-rich and B-rich domains is treated as a
sharp kink (this approximation is popular in theoretical treatments
of wetting \cite{12,13,14,15,16,17,18,19,20}), one might think that
a sharp wetting transition could still be described in terms of the
vanishing of the contact angle $\theta$ as $T \rightarrow T_{\rm w}^-$
(Fig.~\ref{fig2}c). However, it is clear that for a correct treatment
the finite width of the interface needs to be taken into account. Thus,
for finite $D$, the contact angle in Fig.~\ref{fig2}c is ill-defined,
and the transition between the two states depicted in Fig.~\ref{fig2}c is
smooth, because a B-rich nonwet domain may also have a thin A-rich layer
at its surface ($x^{\rm surf}_{\rm A}$, in general, is nonzero). One
should also note that the contact ``line'' is distorted by line tension
effects when it hits the wall, and the line tension of the interface at
the wall would also modify Eq.~(\ref{eq4}) \cite{83,84,85}. The difficulty
of estimating the contact angle in finite geometries is well-known from
studies of nanoscopic droplets \cite{86,87}.

The central conclusion in this subsection is that in the final
equilibrium to which, for times $t\to \infty$ and for small $D$, the
thin film evolves, there is no fundamental difference whether or not we
are above or below the wetting transition temperature, but it matters
whether $T<T_{\rm c}(D)$ or $T>T_{\rm c}(D)$.

\subsection{Bulk phase separation of binary fluid mixtures}
Next, we review our understanding of the kinetics of phase separation in
bulk fluid mixtures which are rendered thermodynamically unstable by a
rapid quench (at $t=0$) into the miscibility gap (see Fig.~\ref{fig2}).
The initial state ($t \leq 0)$ is spatially homogeneous, apart from
small-scale concentration inhomogeneities.  The final equilibrium state
consists of macroscopic domains of the two coexisting phases, with
relative amounts determined by the lever rule.  We are interested in
the evolution from the initial homogeneous state to the final segregated
state.  For quenches below the spinodal curve, the homogeneous system is
unstable and decomposes via the spontaneous growth of long-wavelength
concentration fluctuations \cite{21,22,23,24,25,26} ({\it spinodal
decomposition}).  Understanding the full time evolution from the initial
stages to the late stages of coarsening is a formidable problem, and is
typically accessed by large-scale simulations of coarse-grained models.

Nevertheless, there exist some cases in which simple domain
growth laws can be obtained from analytical considerations
\cite{42,43,44,45,93,94}. The {\it evaporation-condensation mechanism}
of Lifshitz and Slyozov (LS) \cite{93} corresponds to a situation
where a population of droplets of the minority phase (say, A) is in
local equilibrium with the surrounding supersaturated majority phase.
The LS mechanism leads to a growth law (valid for dimensionality $d >
1$) $\ell (t) \propto t^{1/3}$, $t \rightarrow \infty$, where $\ell
(t)$ is the linear dimension of the droplets.

The droplet {\it diffusion-coagulation mechanism} \cite{94} is specific to
fluid mixtures, and is based on Stokes law for the diffusion of droplets,
yielding \cite{94} $\ell(t) \propto ( t/\eta)^{1/d}$.

A faster mechanism of domain growth in fluids was proposed by Siggia
\cite{42}, who studied the coarsening of interconnected domain structures
via the deformation and break-up of tube-like regions, considering a
balance between the surface energy density $\sim \gamma_{\rm AB} /\ell$
and the viscous stress $\sim 6 \pi \eta v_\ell /\ell$ \cite{25}. Thus,
$v_\ell \propto \gamma_{\rm AB} / \eta$ and $\frac{d \ell}{dt} \propto
v_\ell$, or $\ell (t) \propto \frac{\gamma_{\rm AB}}{\eta} t$ in $d=3$.
In $d=2$, the analog of this hydrodynamic mechanism is controversial. San
Miguel {\it et al.} \cite{43} argue that strips ($d=2$ analogs of tubes)
are stable under small perturbations, in contrast to the $d=3$ case. For
critical volume fractions, an interface diffusion mechanism is proposed
which yields $\ell (t) \propto t^{1/2}$, i.e., the same growth law as the
{\it Brownian coalescence} mechanism of droplets in $d=2$ (see above).
On the other hand, Furukawa \cite{44,62} argues for a linear relation
$\ell (t) \propto t$ in $d=2$ as well. However, recently there is growing
evidence \cite{61,62,63} that different characteristic length scales in
$d=2$ may exhibit different growth exponents, suggesting that there is
no simple dynamical scaling of domain growth in $d=2$!

Finally, we remark that the above growth laws do not constitute the true
asymptotic behavior, either in $d=2$ or $d=3$. Rather, these results
only hold for low enough Reynolds numbers \cite{25}.  For $\ell \gg
\ell_{\rm in} = \eta^2/(n \gamma_{\rm AB})$, the so-called inertial
length \cite{25}, one enters a regime where the surface energy density
$\gamma_{\rm AB}/\ell$ is balanced by the kinetic energy density $n
v^2_\ell$. This yields the following growth law for the inertial regime
\cite{25,44}:
\begin{equation} \label{eq13}
\ell(t) \propto \left(\frac{\gamma_{\rm AB}}{n}\right)^{1/3} t^{2/3} \quad ,
\end{equation}
which is valid for both $d=2$ and $d=3$. In $d=2$, evidence for both $\ell
(t) \propto t^{1/2}$ and $\ell (t) \propto t^{2/3}$ has been reported,
but the conditions under which such power laws hold in $d=2$ are still
not clear \cite{50,61,62,63,64,65,66}.

\subsection{Ginzburg-Landau model of surface-directed spinodal
decomposition} 
In this subsection, we briefly discuss a coarse-grained description
of binary mixtures in a thin-film geometry, which can reproduce the
phase diagrams shown in Fig.~\ref{fig2}. This description can also
be used to obtain a model for the kinetics of phase separation in a
confined geometry. Our starting point is a mean-field description of
a binary mixture near its critical point, where one introduces a local
order parameter $\psi(\vec{\rho}, z)$. (Here, $\vec{\rho}$ represents
the coordinates parallel to the walls, and $z$ is the coordinate in the
perpendicular direction, as before.)  The surfaces of the thin film $S_1$
and $S_2$ are located at $z=0$ and $z=D$, respectively.

We denote the order parameter describing the bulk coexistence curve as
$\psi_{\rm b}$, and define $\psi' (\vec{\rho},z)= \psi (\vec{\rho}, z)/
\psi_{\rm b}$. Further, we measure distances $\vec{\rho},z,D$ in units
of $\xi$. Then, the dimensionless free-energy functional of a binary
mixture in a thin-film geometry can be written as a sum of a bulk term
$F_{\rm b}$ and two surface terms \cite{40}, $F [\psi] = F_{\rm b}
[\psi] + F_{S_1} [\psi] + F_{S_2} [\psi]$, where we have dropped the
prime on $\psi'$. Here, we have
\begin{equation} \label{eq15}
F_{\rm b} [\psi] = \int d \vec{\rho} \int\limits^D_0 dz \left[-
\frac{\psi^2}{2} + \frac{\psi^4}{4} + \frac{1}{4} (\vec{\nabla}
\psi)^2 + V(z) \psi\right] \quad .
\end{equation}
The terms $F_{S_1}$ and $F_{S_2}$ are obtained as integrals over the
surfaces $S_1$ and $S_2$:
\begin{equation} \label{eq16}
F_{S_1} = \int\limits_{S_1} d \vec{\rho} \left\{-
\frac{g}{2} [ \psi (\vec{\rho}, 0)]^2 - h_1 \psi (\vec{\rho}, 0) -
\gamma \psi (\vec{\rho}, 0) \frac{\partial \psi}{\partial z}
\bigg|_{z=0} \right\} \quad ,
\end{equation}
and analogously for $F_{S_2}$.

In Eq.~(\ref{eq15}), we have included a $z$-dependent surface potential
which arises due to the surfaces. In our subsequent discussion, we will
consider symmetric power-law potentials:
\begin{equation} \label{eq18}
V(z) =-V_0 \left[(z+1)^{-p} + (D + 1-z)^{-p} \right] \quad ,
\end{equation}
which satisfy $V(z)=V(D-z)$. The potentials are taken to originate behind 
the surfaces so as to avoid singularities at $z=0,D$.

The terms $F_{S_1}$, $F_{S_2}$ represent the surface excess free-energy
contributions due to local effects at the walls, with $g$ and $\gamma$
phenomenological parameters \cite{31,35,70,95}. The dimensionless surface
fields in $F_{S_1}$ and $F_{S_2}$ are $h_1=-V(0)$ and $h_2=-V(D)$,
respectively. The one-sided derivatives appear in $F_{S_1}$ and $F_{S_2}$
due to the absence of neighboring atoms for $z<0$ and $z>D$.

Let us first consider the limit $D \rightarrow \infty$.  For the
long-range surface potential Eq.~(\ref{eq18}), only first-order wetting
transitions are possible \cite{14}. For power-law potentials as in
Eq.~(\ref{eq18}), an approximate theory \cite{35} predicts that the
wetting transition occurs when $2V_0/(p-1) = \gamma_{\rm AB}$.

For finite values of $D$, one can obtain phase diagrams of thin films
(as shown schematically in Fig.~\ref{fig2}) by minimizing the free-energy
functional in Eqs.~(\ref{eq15})-(\ref{eq18}). However, this requires
numerical work \cite{71,78}. We also note that the $\psi^4$-model cannot
describe either the low-temperature region (where complete separation
between A and B occurs), or the non-mean-field critical behavior. 

We next discuss the dynamics of phase separation in thin films.  First,
let us establish the dynamical equations which govern phase separation
in the bulk for the diffusion-driven case.  The local order parameter
$\psi (\vec{r}, t)$ is conserved, and obeys the continuity equation
\cite{21,22,23,24,25,26}:
\begin{equation} \label{eq20}
\frac{\partial}{\partial t} \psi (\vec{r},t) =
- \vec{\nabla} \cdot \vec{J} (\vec{r},t) \quad ,
\end{equation}
The current $\vec{J} (\vec{r},t)$ contains contributions from the local
chemical potential difference $\mu(\vec{r}, t)$, and from statistical
fluctuations, $\vec{\theta} (\vec{r},t)$:
\begin{eqnarray} \label{eq21}
\vec{J} (\vec{r}, t) &=& - \vec\nabla \mu (\vec{r}, t) + \vec{\theta}
(\vec{r}, t) \quad , \nonumber \\
\mu (\vec{r}, t) &=& \frac{\delta F}{\delta \psi (\vec{r}, t)}
\quad .
\end{eqnarray}
Using the $\psi^4$-free-energy functional in Eq.~(\ref{eq15}), we obtain
the dynamical model:
\begin{eqnarray} 
\label{chc}
\frac{\partial} {\partial t} \psi (\vec{r}, t) = \vec{\nabla} \cdot
\left\{ \vec{\nabla} \left[- \psi + \psi^3 - \frac{1}{2} \nabla^2 \psi +
V (z) \right] + \vec{\theta} (\vec{r}, t) \right\}, \quad 0<z<D .
\end{eqnarray}
We assume that the noise $\vec{\theta}$ is a Gaussian white noise, obeying
the relations $\langle \vec{\theta} (\vec{r}, t) \rangle =0$ and $\langle
\theta_i (\vec{r}\;', t') \theta_j (\vec{r}\;'', t'') \rangle = 2 \epsilon
\delta_{ij} \delta (\vec{r}\;'-\vec{r}\;'') \delta (t'-t'')$, where the
indices $i, j$ denote the Cartesian components of vector $\vec{\theta}$.
Note that the time units have been chosen such that the diffusion constant
in Eq.~(\ref{eq21}) is unity.  With respect to the dynamical behavior in
the critical region, Eq.~(\ref{chc}) corresponds to {\it
model B} in the Hohenberg-Halperin classification \cite{99}.  However,
statistical fluctuations are irrelevant for the late stages of spinodal
decomposition \cite{po88}:  The deterministic model obtained by setting
$\epsilon = 0$ in Eqs.~(\ref{eq20})-(\ref{eq21}) also yields the LS
growth law (see Sec.~IIB) in the late stages of domain growth.

One can incorporate hydrodynamic effects, as is appropriate for fluid
mixtures, by including a velocity field $\vec{v} (\vec{r}, t)$ \cite{99},
but this will not be further considered here.

The above models describe coarsening kinetics in the bulk. When one
deals with SDSD in thin films, the model needs to be supplemented by
boundary conditions at the surfaces \cite{28,100}.  The first boundary
condition expresses the physical requirement that the $z$-component of
the flux at the surfaces must vanish:
\begin{equation} \label{eq26}
J_z(\vec{\rho}, 0,t)= \left\{ -\frac{\partial}{\partial z} \left[-
\psi+\psi^3 -\frac{1}{2} \nabla^2 \psi + V(z) \right]+ \theta_z \right
\}_{z=0} = 0 \quad ,
\end{equation}
and similarly for $z=D$. The second boundary condition describes the
evolution of the surface order parameter. Since this quantity is not
conserved, it is described by a relaxational kinetics of model A type
\cite{99}:
\begin{eqnarray} \label{eq27}
\tau_0 \frac{\partial}{\partial t} \psi (\vec{\rho}, 0,t) &=&
- \frac{\delta F}{\delta \psi (\vec{\rho},0,t)} \nonumber\\
&=& h_1 + g \psi (\vec{\rho}, 0, t) + \gamma
\frac{\partial \psi}{\partial z} \bigg|_{z=0} \quad .
\end{eqnarray}
An analogous equation can be written down for the relaxation of
$\psi (\vec{\rho}, D, t)$. Here, $\tau_0$ sets the time-scale of this
nonconserved kinetics. Since $\psi(\vec{\rho}, 0, t)$ relaxes much faster
than the order parameter in the bulk, it is reasonable to set $\tau_0=0$
\cite{35}. Then, the dynamics as well as the statics is controlled by
two surface parameters, $h_1/\gamma$ and $g/\gamma$. During the early
stages of SDSD, the fast relaxation of the order parameter at the
surfaces provides a boundary condition for the phase of concentration
waves that grow in the thin film. In the bulk, the random orientations
and phases of these growing waves do not yield a systematic evolution of
the average order parameter. However, the surface-directed concentration
waves add up to give an average oscillatory concentration profile near
the surfaces of a thin film \cite{27,28,29,30,31,32,33,34,35,39,40}.

Our early work on SDSD in thin films \cite{39} omitted both hydrodynamic
interactions and the noise in Eq.~(\ref{eq21}), and focused on the $d=2$
case. In recent work \cite{40}, we have studied the $d=3$ case using the
GL model in Eq.~(\ref{chc}) with the noise term, in conjunction with the
boundary conditions in Eqs.~(\ref{eq26})-(\ref{eq27}). In Sec.~IV, we will
compare our MD results with results from this study. The details of the GL
simulation are as follows. We implemented an Euler-discretized version of
Eqs.~(\ref{chc}),~(\ref{eq26})-(\ref{eq27}) on an $L \times L \times D$
lattice. The discretization mesh sizes were $\Delta x = 1$ and $\Delta t
= 0.02$. The surface potential was of the form in Eq.~(\ref{eq18}) with
$p=3$, which corresponds to a non-retarded van der Waals' interaction
between the surfaces and a particle in $d=3$. The parameter values
were $g=-0.4,\gamma=0.4$, and $V_0 = 0.325$ for $D=5$ and $V_0 = 0.11$
for $D=10$, corresponding to a partially wet surface in equilibrium.
We stress that, for a fluid mixture, the above diffusive model is relevant
during the early stages of phase separation \cite{21,22,23,24,25,26},
but is not expected to yield useful results for the intermediate and
late stages of domain growth.

\section{Model and Molecular Dynamics Methods}
For our MD study, we consider a fluid of point particles located in
continuous space in a box of volume $L \times L \times D$. We apply
periodic boundary conditions in the $x$ and $y$ directions, while
impenetrable walls are present at $z=0$ and $z=D$. These walls give rise
to an integrated LJ potential ($\alpha$ = A,B):
\begin{equation} \label{eq28}
u_{\rm w} (z)= \frac{2 \pi n \sigma^3}{3} \epsilon_{\rm w}
\left[\frac{2}{15} \left(\frac{\sigma}{z'}\right)^9 - \delta_\alpha
\left(\frac{\sigma}{z'}\right)^3 \right] \quad ,
\end{equation}
where $n$ is the reference density of the corresponding bulk fluid
\cite{36,37,38}, $\sigma$ is the LJ diameter of the particles, and
$\epsilon_{\rm w}$ is an energy scale for the strength of the wall
potentials. Further, $\delta_{\rm A}=1$ and $\delta_{\rm B}=0$, so
A-particles are attracted by the walls while B-particles are not.
The coordinate $z'=z + \sigma/2$ for the wall at $z=0$, and $z'=D+
\sigma/2 -z$ for the wall at $z=D$. Therefore, the singularities of
$u_{\rm w}(z)$ do not occur within the range $0\leq z \leq D$ but rather
at $z= - \sigma/2$ and $z=D + \sigma/2$, respectively.

The particles in the system interact with LJ potentials:
\begin{equation} \label{eq29}
u(r_{ij})= 4 \epsilon_{\alpha \beta} \left[ \left(\frac{\sigma_{\alpha
\beta}}{r_{ij}}\right)^{12} - \left(\frac{\sigma_{\alpha \beta}}{r_{ij}}
\right)^6 \right], \quad \quad r_{ij}=|\vec{r}_i-\vec{r}_j|,
\end{equation}
where $\alpha, \beta = {\rm A}, {\rm B}$. The LJ-parameters
$(\epsilon_{\alpha \beta}, \sigma_{\alpha \beta})$ are chosen as follows:
\ba \label{eq30}
&& \sigma_{\rm AA} =\sigma_{\rm AB}= \sigma_{\rm BB}= \sigma,
\nonumber \\
&& \epsilon_{\rm AA}=\epsilon_{\rm BB}=\epsilon, \quad
\epsilon_{\rm AB}=\frac{\epsilon}{2} \quad .
\ea
The units of length, temperature, and energy are chosen such that
$\sigma=1$, $\epsilon=1$, $k_B=1$. The masses of the particles are
chosen to be equal, $m_{\rm A}=m_{\rm B}=m=1$. Thus, the MD time unit
\cite{101,102,103}:
\be
t_0 = \left(\frac{m \sigma^2}{48 \epsilon}\right)^{1/2} =
\frac{1}{\sqrt{48}} ,
\ee
becomes a dimensionless number. To speed up the calculations, the LJ
potential is truncated at $r_{ij}=2.5\, \sigma$ and shifted to zero
there, as usual \cite{101}. To ensure that our study of fluid-fluid phase
separation is not affected by other phase transitions (e.g., liquid-gas
or liquid-solid transitions), we work with bulk density $n=1.0$ and focus
on temperatures $T>1.0$.  In principle, in thin films one could have
a wall-induced crystallization at temperatures above the bulk melting
temperature: however, we have not seen any evidence for such an effect in
our model.  In our previous work on the bulk behavior of the same model
\cite{36,37,38}, we found that the critical temperature for bulk phase
separation is $T_{\rm c} \simeq 1.638$. Here, we present results from
simulations of quenching experiments to $T=1.1$. At this temperature,
the bulk phase separation is essentially complete.  In addition,
the bulk correlation length $\xi \simeq 1$ (i.e., one LJ diameter)
within the relative accuracy of about 5\% to which it can be determined
\cite{36}. Further, material parameters which enter the theories reviewed
in Sec. II (e.g., the interfacial tension $\gamma_{\rm AB}$, the shear
viscosity $\eta$) are explicitly known as well. The appropriate values
are $\gamma_{AB} \simeq 0.9$ (see Ref.~\cite{38}) and $\eta \simeq 7$
(see Ref.~\cite{36}).  (Recall that these quantities are measured in LJ
units and hence are dimensionless.) The availability of most material
parameters for our system is a distinct advantage of our atomistic model
in comparison to coarse-grained models, where it is often unclear what
ranges of effective parameters correspond to physically reasonable
choices.

The strength of the wall-particle interaction is taken as $\epsilon_{\rm
w} =0.005$, which corresponds to partially wet walls at $T=1.1$. To
find the precise location of the wetting transition for our model
would require a major computational effort, and this has not been
attempted.  Recall that one expects $T_{\rm w} \rightarrow T_{\rm c}$
when $\epsilon_{\rm w} \rightarrow 0$ -- this was the motivation for
choosing a rather small value of $\epsilon_{\rm w}$ in our study.
However, due to the special choices made [such as Eq.~(\ref{eq30})], for
the sake of simplicity, it would be premature to try to explicitly
relate our model to a specific real system.

The lateral size $L$ of the simulated systems must be large enough that
the laterally inhomogeneous structures that form during segregation
are not affected by finite-size effects. Therefore, we chose $L=128$
for the thinnest film in our study ($D=5$) and $L=64$ for the thicker
ones ($D=10,20$). As the confining potentials diverge at $z=-\sigma/2$
and $z=D+ \sigma/2$ ($\sigma=1$), the volume in which particles can be
is $V=L^2 (D+1)$. We will report results from three sets of simulations,
with $D=5,L=128$ ($N=98304$ particles); $D=10,L=64$ ($N=45056$); and
$D=20,L=64$ ($N=86016$). Thus, the particle density is $n=1$ in all
these cases. For the range of times studied here ($t\leq8000$), test runs
with other linear dimensions showed that our choices of $L$ are large
enough to eliminate finite size effects, within the limits of our
statistical accuracy. Of course, for a study on larger time scales
also larger system sizes would be required!

The initial states of the simulations need to be carefully prepared.
We equilibrated a fluid of $N$ particles (with $N_{\rm A} = N_{\rm B}
= N/2$) in the specified volume at a very high temperature ($T=5$),
with periodic boundary conditions in all directions. The equilibration
time was $10^5$ MD time steps. At this high temperature, only very weak
chemical correlations develop among the particles. We use the standard
velocity Verlet algorithm \cite{101,102,103} with a time step of 0.02, and
apply the Nos\'e-Hoover algorithm \cite{101,102,103} for thermalization.

At time $t=0$, the wall potentials are introduced, and the temperature
is quenched to $T=1.1$. This is done by rescaling the velocities,
and by setting the temperature of the Nos\'e-Hoover thermostat to the
new temperature. Of course, in a real experiment, the temperature of a
fluid confined in a small slit pore would be controlled via the thermal
energy of the solid walls forming the pore.  Therefore, no instantaneous
quench (on picosecond or nanosecond time-scales) is possible. However,
the structure formation occurring in a binary fluid with a finite quench
rate is a complication that we disregard here.

\begin{figure}
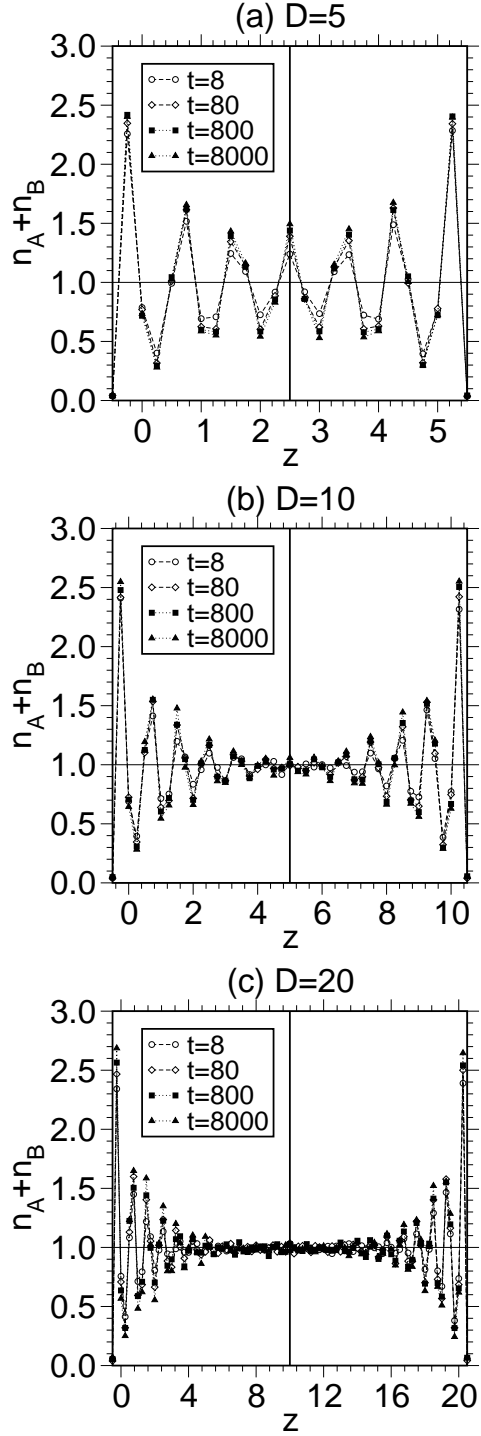
 
\centering
\includegraphics*[width=.38\textwidth]{fig3a.eps} \\
\vskip0.2cm
\includegraphics*[width=.38\textwidth]{fig3b.eps} \\
\vskip0.2cm
\includegraphics*[width=.38\textwidth]{fig3c.eps} \\
\caption{\label{fig3}
Profiles of the total density $n(z)=n_{\rm A}(z) + n_{\rm B}(z)$
vs. $z$ for (a) $D=5$, (b) $D=10$, and (c) $D=20$. We show data
for four different times $t$ (in units of $t_0$), as
indicated. The vertical line in each frame indicates the
mid-plane of the film at $z=D/2$. The wall potentials
diverge at $z=-1/2$ and $z=D+1/2$ [see Eq.~(\ref{eq28})], so the
particles can range over a distance $(D+1)$ in the
$z$-direction. All lengths are measured in units of $\sigma$,
and hence are dimensionless.}
\end{figure}

It is also relevant to discuss our procedure of introducing the walls
together with the quench at time $t=0$. In this case, the fluid is
translationally invariant for $t<0$, but loses this invariance in the
$z$-direction for $t>0$.  However, we have found that the typical
oscillatory density profiles near the walls (``layering'') already
develop during the first few MD time steps after the quench -- see
Fig.~\ref{fig3}. For $D=5$, we recognize 7 well-developed layers and there
is no region of constant density in such an ultra-thin film. However,
for $D=10$, the region from $z \simeq 4$ to $z \simeq 6$ has an almost
constant density $n(z)=n_{\rm A} (z) + n_{\rm B} (z) \simeq n=1$. For
the $D=20$ case, this constant density region covers about half of the
film thickness, extending from $z \simeq 5$ to $z \simeq 15$. Note that
the layer distance in Fig.~\ref{fig3} is slightly less than $\sigma$,
although $\sigma$ coincides with the position of the first peak of the
radial distribution functions in the bulk \cite{36}.
We introduce the walls together with the quench at time $t=0$ to make
the initial state of the quench (random distribution of $A$ and $B$
particles everywhere in the system, also close to the wall) comparable
to that of the Ginzburg-Landau model (where we quench from a state at
``infinite temperature'').

It has been emphasized \cite{102} that MD simulations constitute a method
to explore hydrodynamic phenomena, and are competitive with coarse-grained
methods. In principle, this is only true for a microcanonical MD in the
NVE ensemble where the energy $E$ is strictly conserved. However, here we
use the Nos\'e-Hoover thermostat \cite{101,102,103}, i.e., we integrate
the following equations of motion for the coordinates of the particles
$\vec{r}_i(t)$ (with ${\dot{\vec{r}}}_i = d\vec{r}_i/dt =\vec{v}_i$):
\begin{equation} \label{eq31}
{\ddot{\vec{r}}}_i(t)=\frac{\vec{f}_i}{m_i} -\zeta(t) \vec{r}_i ,
\end{equation}
\begin{equation} \label{eq32}
\dot{\zeta} (t)=\frac{1}{Q} \left(\sum\limits_{i=1}^N m_i {\vec{v}_i}^{~2}
- 3 Nk_BT\right) \quad .
\end{equation}
Here, $Q$ is the fictitious mass of the thermostat, which was set to
$Q=100$. In the limit $Q = \infty$, we have $\zeta (t) = 0$ and then
we recover the strict conservation of energy and momentum, on which the
equations of hydrodynamics are based. For finite $Q$, we have $\langle
\zeta (t) \rangle=0$. However, the fluctuating damping term disturbs
hydrodynamics slightly.  For this reason, in our earlier study of
isothermal transport coefficients in the bulk \cite{36,37}, we have used
strictly microcanonical runs. However, the ensemble of initial states was
generated by Monte Carlo simulations in the semi-grand-canonical ensemble,
ensuring thus a strict validity of all conservation laws in conjunction
with averaging in the NVT ensemble. In the context of a thin fluid film
confined in a slit between solid walls formed from vibrating atoms,
neither momentum nor energy (of the fluid film) are conserved, and the
walls do act as a thermostat. Of course, in a more realistic model,
this thermostatting action of the walls applies only to those fluid
particles which are close to one of the walls, and not on particles
near the center of the film (which are only thermostatted indirectly
via heat conduction). In situations far from local equilibrium (such as
strongly sheared fluids \cite{103,104}), there is indeed a noticeable
difference between the effects of wall thermostats and the Nos\'e-Hoover
thermostat. However, we do not expect any such problem here, since the
time-scales for domain coarsening are much larger than the time-scales
associated with heat conduction.

In simulations of domain growth, one encounters the problem of large
statistical fluctuations, and quantities such as the {\it equal-time
correlation function} $C(\vec{r},t)$ exhibit lack of self-averaging
\cite{105}.  [Unlike the equilibrium case, $C(\vec{r},t)$ explicitly
depends on the time $t$ after the quench.] Such quantities can only
be sampled if a number of independent runs are performed and averaged
over. All statistical quantities presented here are obtained as averages
over three independent runs.

\section{Numerical Results}
Let us begin with a discussion of the laterally averaged order parameter
profiles, $\psi_{\rm av} (z,t)$ vs.~$z$ (see Fig.~\ref{fig4}). These
are obtained from our MD simulations by averaging individual profiles
for $\psi (x,y,z,t)$ vs. $z$ along the $x$ and $y$ directions, and
further averaging over independent runs. The morphology of these SDSD
profiles consists of an A-rich wetting layer at the surface, followed
by a depletion layer in A, etc. One can see that the order parameter at
the surface has already increased to a rather large value at early times,
due to the preferential attraction of the A-particles to the walls. These
A-particles are removed from the adjacent regions in the interior of the
film, resulting in local minima in $\psi_{\rm av} (z,t)$ (A-depletion
layers). As time proceeds, these minima move towards the center, and
also become more pronounced. In the $D=5$ case, the SDSD waves coalesce
rapidly and only a single minimum in the center is left by $t=640$. In
the $D=10$ case, distinct SDSD waves are visible till $t \simeq 1000$. At
$t=2000$ (see Fig.~\ref{fig4}d), the waves have merged to give a layered
structure with a single minimum.  In both cases, the layered structure is
transient and breaks up into a columnar structure which coarsens laterally
(see Fig.~\ref{fig1}).  Of course, even in this asymptotic state, the
walls remain A-rich and the film center is B-rich.  This behavior is
reminiscent of the SDSD profiles seen in GL studies of this problem
\cite{28,31,39,40}, and corresponding experiments \cite{30}. In the
present MD simulations, only a single depletion minimum is observed
near each wall -- statistical fluctuations of the local position of the
boundaries between the depletion layers and adjacent enrichment layers
wipe out any further systematic variation of the concentration profiles.

It is also interesting to examine the evolution of the local
order parameter $\psi_{\rm av} (0,t)$ at the surface $z=0$ (see
Fig.~\ref{fig5}a) or $z=D$ (which is analogous to Fig.~\ref{fig5}a).
Note that we have averaged the MD data over a layer of thickness $\Delta
z=1$ to estimate $\psi_{\rm av} (0,t)$ (whereas for the calculation
of $\psi_{\rm av} (z,t)$ in Fig.~\ref{fig4} $\Delta z=0.25$ was used).
The quantity $\psi_{\rm av} (0,t)$ rises rapidly, and reaches a maximum
at about two decades, before it starts decreasing.  The rapid rise is
expected from the phenomenological theory (see Sec.~II C).  Of course,
due to the lack of conservation of the local order parameter adjacent
to the walls, there is an immediate response to the surface potential
at the wall.  For both $D=5$ and $D=10$, even runs up to $t=2\cdot
10^4$ do not suffice to estimate the final values of $\psi_{\rm av}
(0,t)$ clearly. This non-monotonic relaxation is a consequence of the
structural rearrangement of the concentration inhomogeneities in the
films. Figure~\ref{fig5}b shows corresponding data for $\psi_{\rm av}
(0,t)$ vs. $t$ from our recent simulations, using the GL model described
in Sec.~II.C \cite{40}. (Note that a logarithmic time-axis is chosen in
Fig.~\ref{fig5}b.) The behavior of the GL data is qualitatively similar
to that of the MD results. However, a pronounced intermediate plateau is
formed in the GL case, whereas the MD data only show a maximum -- see the
inset of Fig.~\ref{fig5}b, which plots the data from Fig.~\ref{fig5}a
on a logarithmic time-scale.  The reason for this difference lies in
the formation of a long-lived metastable layered state with pronounced
A-rich layers in the GL case, which is not observed in the MD simulations.

One way to further elucidate the morphological evolution in the MD
simulations is to look at snapshot pictures of the concentration in
cross-section planes through the films (see Fig.~\ref{fig6}). At $t=80$,
the distribution of the particles is rather random, but by $t=800$ the
existence of domain structures is fairly evident. This interconnected
structure coarsens ($t=1600$), and ultimately breaks up into compact
domains that connect the enrichment layers at both walls. (In the snapshot
for $D=10$, at $t=8000$, only the enrichment layers are seen, but this
observation is accidental. The slice shown cuts through a region free
of columnar A-rich domains over the lateral scale $L=64$, while other
slices parallel to the one shown do cut through such domains. But we
include this example in order to emphasize that strong fluctuations
occur, not only from one run to the next run, but also in the course
of the time evolution of individual runs. Thus, individual snapshot pictures
give qualitative insight only.)

Thus, Fig.~\ref{fig6} already provides evidence of the simultaneous
presence of surface enrichment layers and lateral phase separation. A
clear picture of lateral phase separation is obtained if we examine the
concentration distribution in slices (of width $\sigma$) centered at the
$L \times L$ mid-plane at $z=D/2$ (see Fig.~\ref{fig7}). These pictures
resemble snapshot pictures of 2-$d$ spinodal decomposition, though for an
off-critical composition. Of course, the concentration is conserved in a
strictly 2-$d$ system, whereas the concentration is not conserved in the
shown $L \times L \times \sigma$ slice. As a matter of fact, it decreases
systematically with increasing time, due to the progressive formation
of A-rich surface enrichment layers. This is particularly evident in
the late-time snapshots ($t=8000$) for $D=10$ and $D=20$, respectively.

In order to quantitatively characterize the lateral phase separation,
we introduce the layer-wise correlation function:
\begin{equation} \label{eq33}
C(\rho,z,t) = \langle \psi (0, z, t) \psi (\vec{\rho}, z,
t)\rangle - \langle \psi (0,z,t) \rangle \langle \psi (\vec{\rho},
z, t) \rangle .
\end{equation}
We also define a layer-wise length scale $\ell(z,t)$ from the decay
of this function with lateral distance $\rho$:
\begin{equation} \label{eq34}
C(\rho = \ell,z,t) = \frac{1}{2} C(0,z,t) \quad .
\end{equation}
In Fig.~\ref{fig8}, we plot the scaled layer-wise correlation function,
$C(\rho,z,t)/C(0,z,t)$ vs. $\rho/\ell(z,t)$ at $t=8000$, for $D=5,10,20$
and different values of $z$. The surface ($z=0$) is strongly enriched in
the preferred component A -- see Fig.~\ref{fig4} and Fig.~\ref{fig5}.
The corresponding correlation function measures small fluctuations
about a strongly off-critical background. The correlation functions
for the inner regions of the film do not scale either. (If there were
scaling, all the data sets in Figs.~\ref{fig8}a-c would superimpose,
as they do when similar plots are made in studies of bulk spinodal
decomposition.)  This is because the correlation function is a function
of the off-criticality \cite{op87,sp88}, and different values of $z$
are characterized by different average compositions -- see the depth
profiles for $t=8000$ in Fig.~\ref{fig4}b (for $D=5$) and Fig.~\ref{fig4}d
(for $D=10$).

In Fig.~\ref{fig9}, we show the scaled correlation function in the film
center for $D=5,10,20$ at different times. Again, there is no scaling
of the data sets. This lack of scaling is expected, however, since the
average volume fraction of A in the central region changes with time
-- see Fig.~\ref{fig4}.  For the case of $D=5$, there is a reasonable
superposition of the curves for $t=4000$ and $t=8000$. This is consistent
with the observation that the average concentration at the center remains
approximately unchanged over this time-regime -- see Fig.~\ref{fig4}b.
In the asymptotic regime, the system evolves via the lateral coarsening
of columnar domains. Therefore, we expect the depth profiles $\psi_{\rm
av}(z,t)$ vs.~$z$ (as in Fig.~\ref{fig4}) to become independent of time
at sufficiently large times. In this regime, we will recover dynamical
scaling for the layer-wise correlation functions.

In Fig.~\ref{fig10}, we plot the layer-wise length scale [$\ell(z,t)$
vs. $t$] for $D=5,10,20$ on a log-log plot, in order to check for
possible power laws.  At early times, no well-defined power law can
be identified at all.  This is not surprising as one does not expect
a universal growth law to apply when the length scale is of the same
order as the inter-particle distance. The gradual increase of the
slope of $d[\ln \ell (t)]/ d(\ln t)$ is consistent with data from
experimental and simulational studies of spinodal decomposition in the
bulk \cite{21,22,23,24,25,46,47,48,49,50}.  Surprisingly, at later times,
the MD data appear to be compatible with a power law with an effective
exponent $\simeq 2/3$.  We do not see evidence for any of the other growth
laws discussed in the context of fluids (see Sec.~IIB)
over an extended period of time. One might have expected that the LS
evaporation-condensation mechanism or the droplet
diffusion-coagulation mechanism would dominate
over some time-range, but this is not the case. As regards the Siggia
tube-coarsening mechanism, the interconnected domain
structures break up so early that hydrodynamic mechanisms can hardly
become operative.

It would be premature to claim that the log-log plots in Fig.~\ref{fig10}
are evidence that the inertial mechanism [Eq.~(\ref{eq13})] has been
seen. According to theory, this mechanism should be visible only if
the length scale $\ell(t) \gg \ell_{\rm in} = \eta^2/(n\gamma_{\rm AB})$.
Fortunately, the material parameters which determine $\ell_{\rm in}$
are known for our model, as emphasized in Sec. III. Putting the numbers
in, we estimate that $\ell_{\rm in} \sim O(10^2)$!  Such large values
of $\ell_{\rm in}$ are compatible with studies of the late stages of
domain growth in $d=3$ using the lattice Boltzmann method \cite{53}.

One might then conclude that the effective power law $\ell(t) \propto
t^{2/3}$ seen in Fig.~\ref{fig10} is only a transient phenomenon,
and the power laws that are expected in this case (see Sec.~IIB)
will come into play at later times. An alternative
possibility is that novel growth laws arise due to the interplay of
wetting kinetics and lateral phase separation in the ``bulk'' of the film.
However, we note that our results have a striking qualitative similarity
to the Brownian-dynamics results of Farrell and Valls \cite{50}. These
authors studied phase separation in strictly 2-$d$ fluid mixtures,
and found a rapid crossover to a power law $\ell (t) \propto t^\phi$
with $\phi \simeq 2/3$.

Obviously, more work with both simulations and theory is needed to
resolve the nature of applicable growth laws.  However, this cannot
be done by simply running our simulations longer. This is because the
condition $\ell(z,t) \ll L$ is needed to ensure that finite-size effects
in the lateral direction are negligible.  Further, the condition $\ell
(z,t) \ll L$ is also needed to provide a reasonable self-averaging of
$C(\rho,z,t)$ [defined in Eq.~(\ref{eq33})]. One can divide the system
laterally into independent blocks of linear size $\ell(z,t)$ to judge the
error in the estimation of $C(\rho, z,t)$. Therefore, the relative error
is less for $D=5$ (where $L=128$) than for $D=10$ and 20 (where $L=64$),
and it increases when $\ell(z,t)$ increases.  Thus, the irregularities
in $\ell(z,t)$ for $D=10,20$ when $t \geq 1000$ are probably due to
insufficient statistics (as only three independent runs were made).

\section{Summary and Discussion}
Let us conclude this paper with a summary and discussion of
our results. Here, we have presented comprehensive results from
molecular dynamics (MD) simulations of {\it surface-directed spinodal
decomposition} (SDSD). We have used a simple model system, namely a
symmetric binary Lennard-Jones (LJ) mixture, confined between identical
flat and structureless parallel walls which preferentially attract the
A-particles. Only very thin films are accessible -- the distance between
the origins of the wall potentials was $(D+1)=6,11,21$ in units of the
LJ parameter.  Further, the finite size of the lateral linear dimension
$L$ ($L=128$ for $D=5$, and $L=64$ for $D=10,20$) constrains our work
to the early and intermediate stages of domain growth. In this regime,
the characteristic length scale of lateral phase separation $\ell(z,t)$
has grown by approximately one decade. Note that we have also restricted
attention to deep quenches, much below the critical temperature of phase
separation, but above the triple-point temperature, so crystallization
is not an issue in our study. At the chosen temperature ($T=1.1$, i.e.,
$T/T_{\rm c} \simeq 0.67$), the bulk phase separation occurs between
almost pure A and B fluids. The interfaces are locally sharp (with a
correlation length $\xi \sim 1$), and the time-scale for structural
relaxation in the fluids is manageable for MD work. The shear viscosity
has been estimated previously \cite{37} to be $\eta \simeq 7$ at $T=1.1$,
in the standard LJ units. Thus, the advantages of the present approach
are as follows: (a) all material parameters of the model are explicitly
known; (b) at small scales, a qualitatively reasonable description of
fluid structure is ensured; and (c) long-range hydrodynamic interactions
(resulting from the conservation laws in fluid dynamics) are automatically
included, although only a short-range LJ potential is chosen to model
the interaction among the point particles.

\begin{figure} 
\centering
\includegraphics*[width=0.9\textwidth]{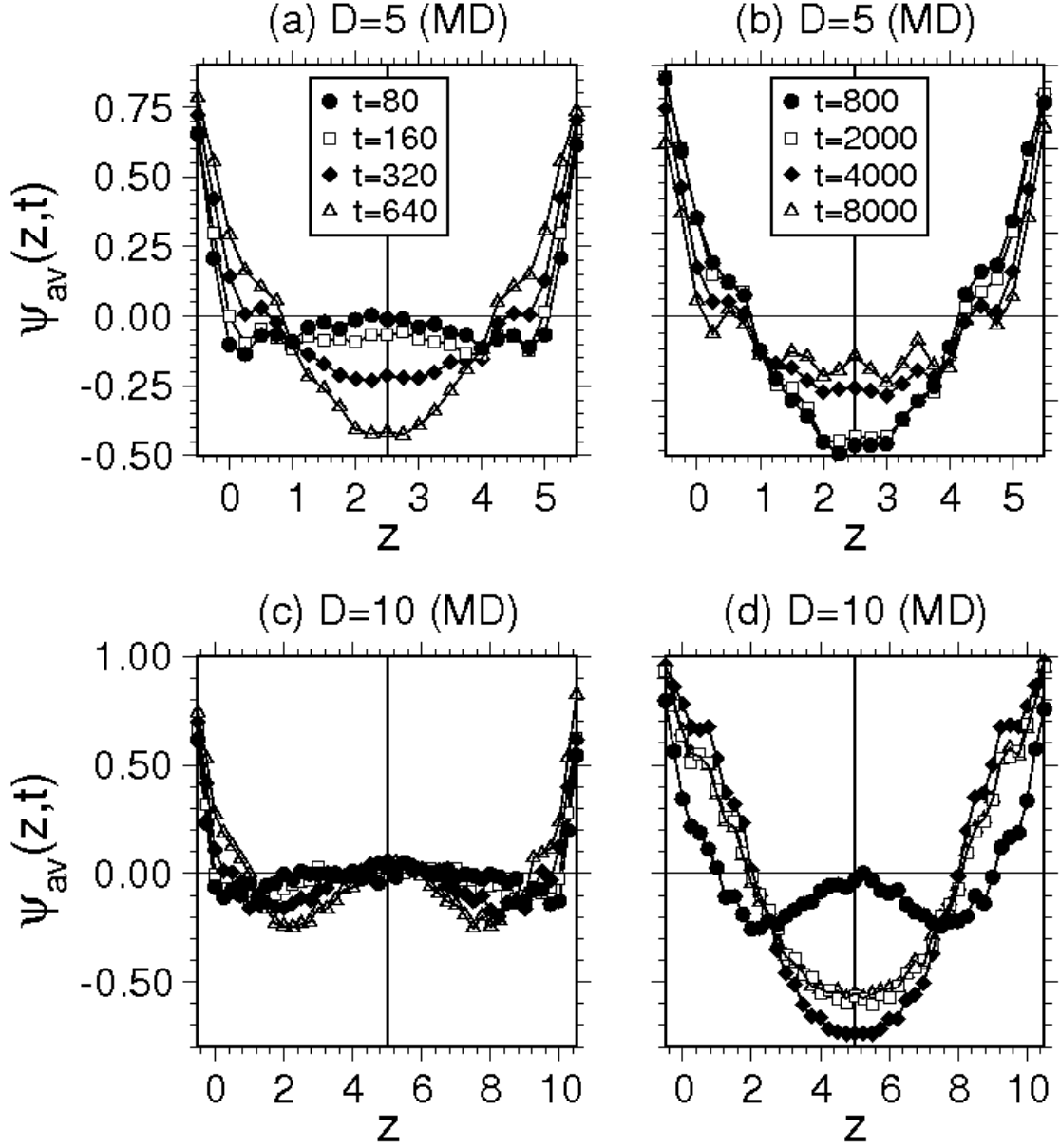}
\caption{\label{fig4}
Laterally averaged order parameter profiles, $\psi_{\rm av} (z,t)$ vs. $z$,
for films of thickness (a) $D=5$ at early times ($t=80,160,320,640$);
(b) $D=5$ at late times ($t=800,2000,4000,8000$); (c) $D=10$ at early
times ($t=80,160,320,640$); (d) $D=10$ at late times
($t=800,2000,4000,8000$). The symbol usage is the same for (a),(c)
as well as for (b),(d).}
\end{figure}

\begin{figure}
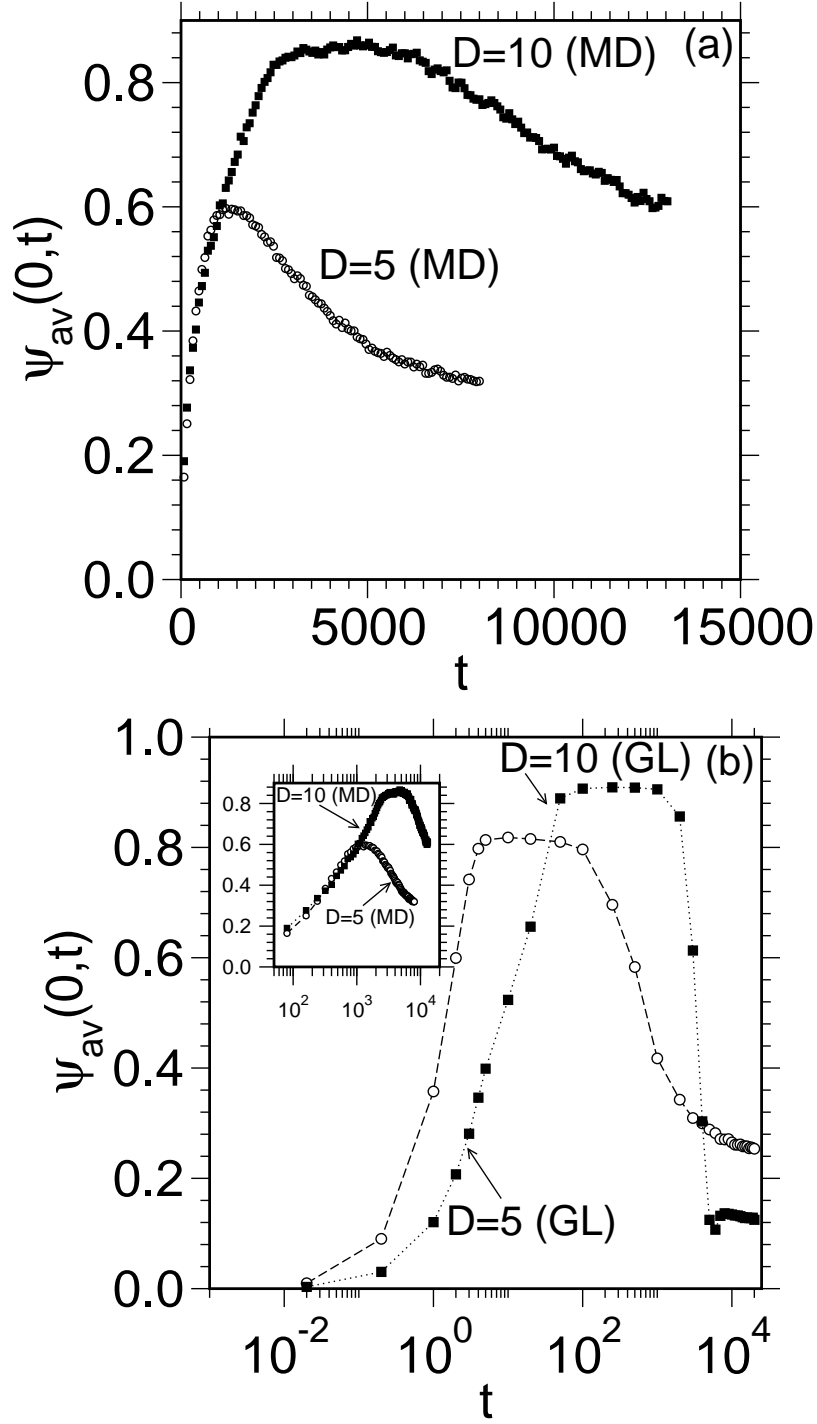
 
\centering
\includegraphics*[width=0.65\textwidth]{fig5a.eps}
\vskip0.25cm
\includegraphics*[width=0.6\textwidth]{fig5b.eps}
\caption{\label{fig5}
(a) Time-dependence of the local order parameter $\psi_{\rm av} (0,t)$
at the surface $z=0$. We show data for $D=5$ and $D=10$. Note that we have
averaged the MD data over a layer of thickness $\Delta z=1$ to estimate
$\psi_{\rm av} (0,t)$. (b) Time-dependence
of $\psi_{\rm av} (0,t)$, obtained from the Ginzburg-Landau (GL) simulations
described in Sec.~II.C \cite{40}. The GL data was obtained as an average
over 5 independent runs with $L=256$.  We plot the data on a log-linear
scale. The inset shows the MD data from (a), also on a log-linear scale.}
\end{figure}

\begin{figure} 
\centering
\includegraphics*[width=0.3\textwidth]{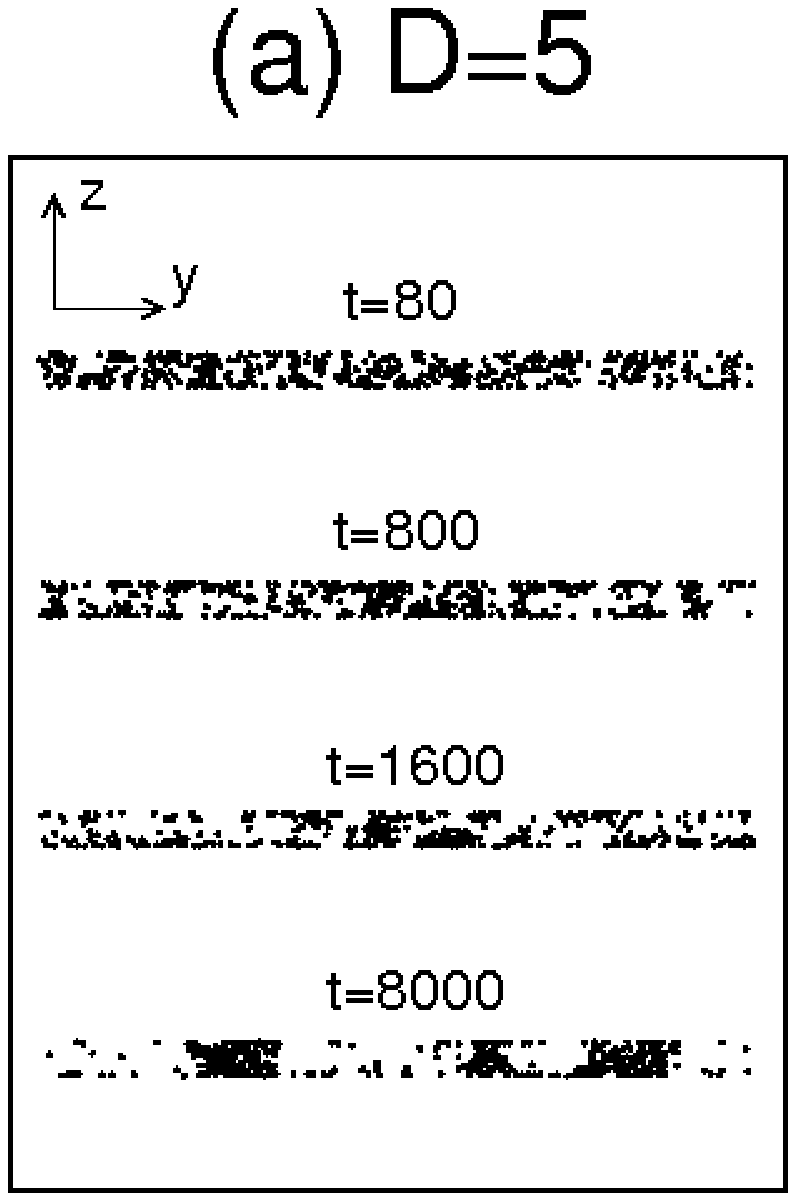}
\includegraphics*[width=0.3\textwidth]{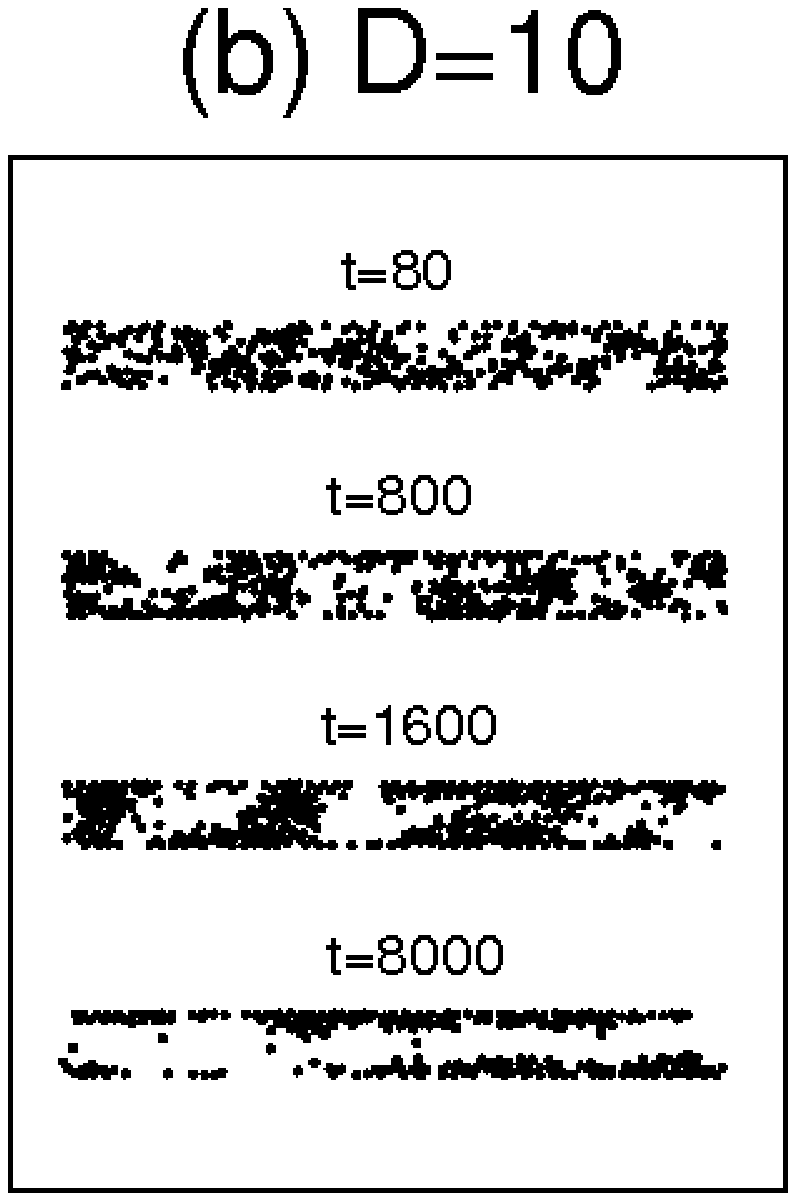}
\includegraphics*[width=0.3\textwidth]{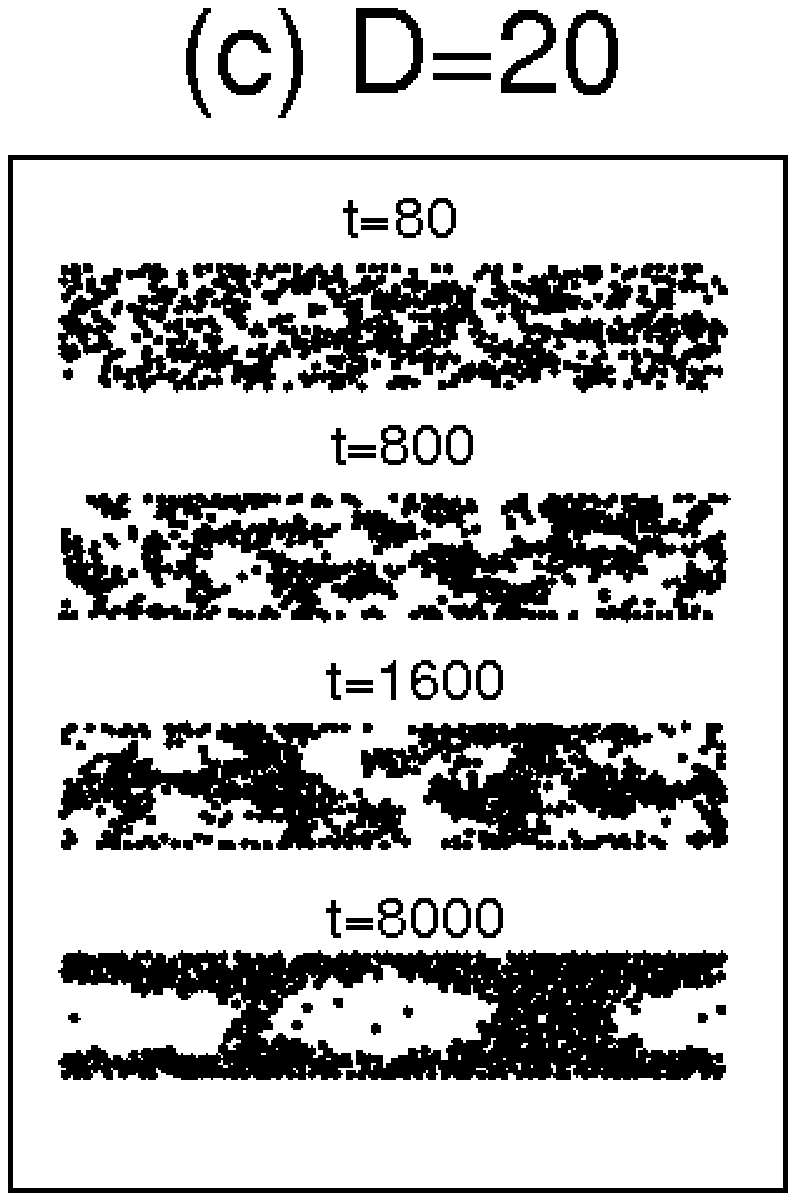}
\caption{\label{fig6}
Snapshots of the concentration distribution in cross-section
slices (of linear dimensions $\sigma \times L \times D$)
through the films, for (a) $D=5$, (b) $D=10$, and (c) $D=20$. The
cross-section was centered at $x=L/2$. The A-particles
are marked black, while the B-particles are not shown. These
pictures correspond to the times $t=80,800,1600,8000$ in all
cases.}
\end{figure}

\begin{figure} 
\centering
\includegraphics*[width=0.45\textwidth]{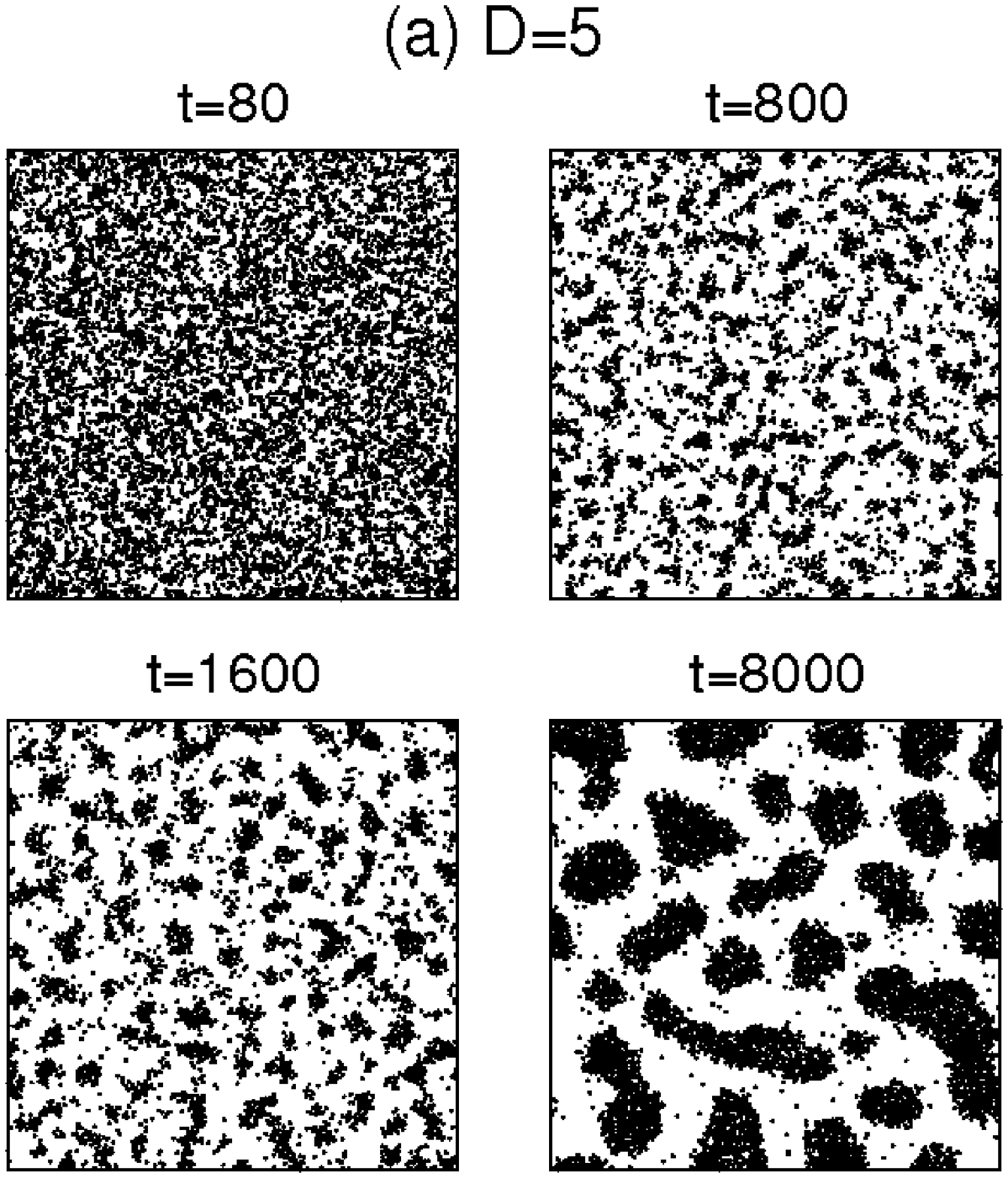}
\includegraphics*[width=0.45\textwidth]{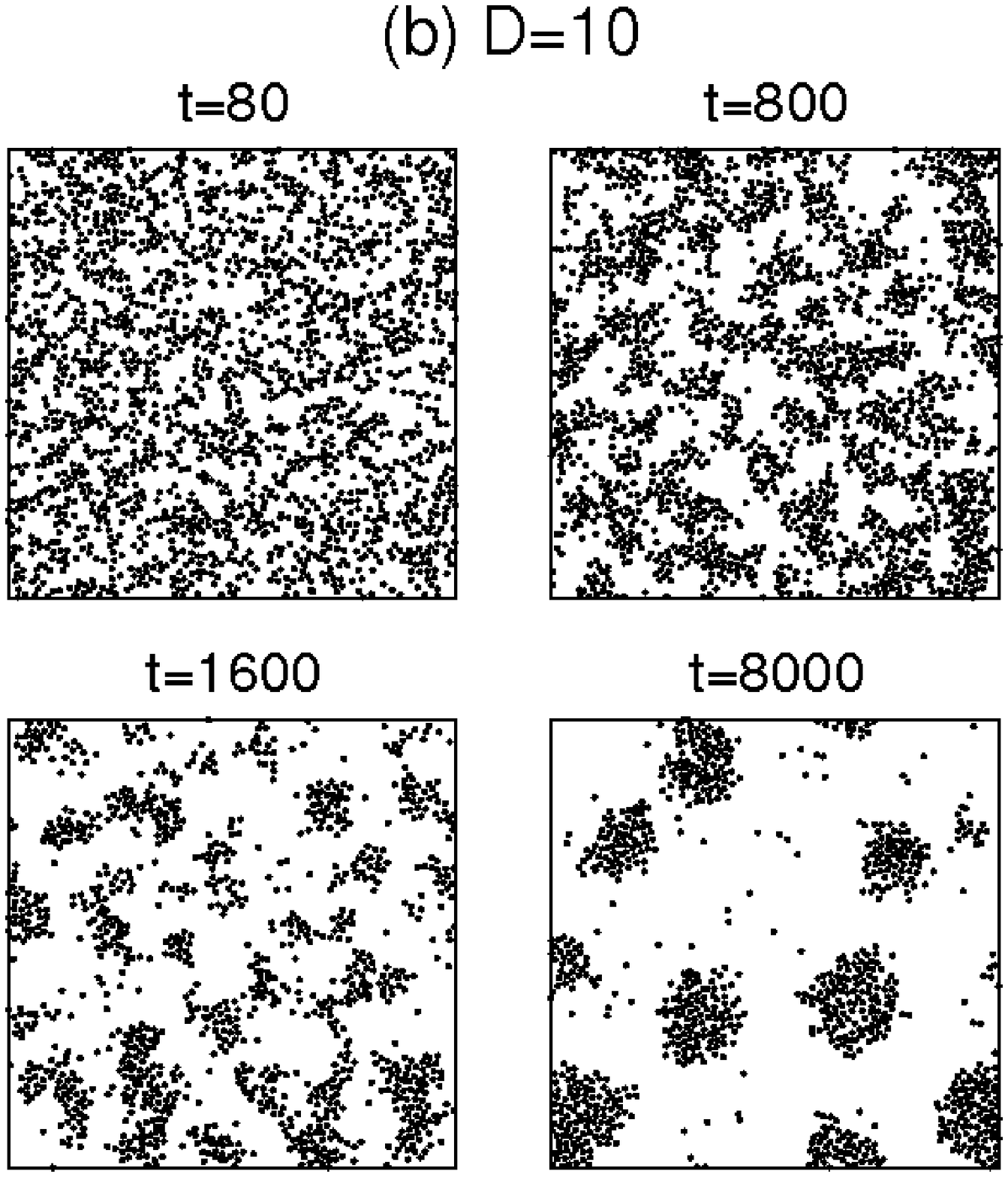}
\vskip0.5cm
\includegraphics*[width=0.45\textwidth]{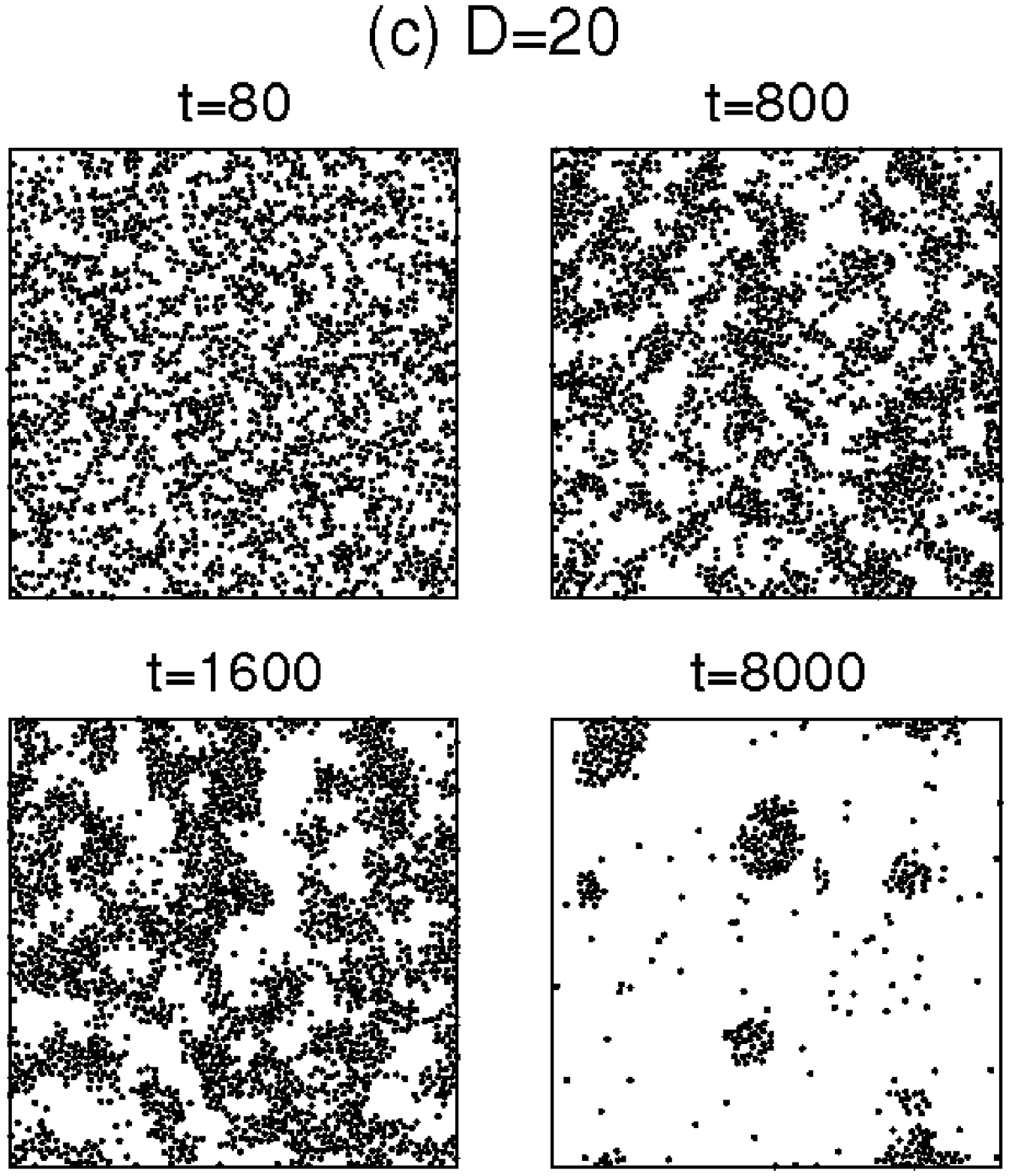}
\caption{\label{fig7}
Snapshots of the concentration distribution in cross-section
slices (of linear dimensions $L \times L \times \sigma$) centered
at the plane $z=D/2$ of the films. We show pictures for
(a) $D=5$, (b) $D=10$, and (c) $D=20$. The A-particles
are marked black, while the B-particles are not shown.
Note that $L=128$ for $D=5$, but $L=64$ for $D=10,20$.}
\end{figure}

\begin{figure}
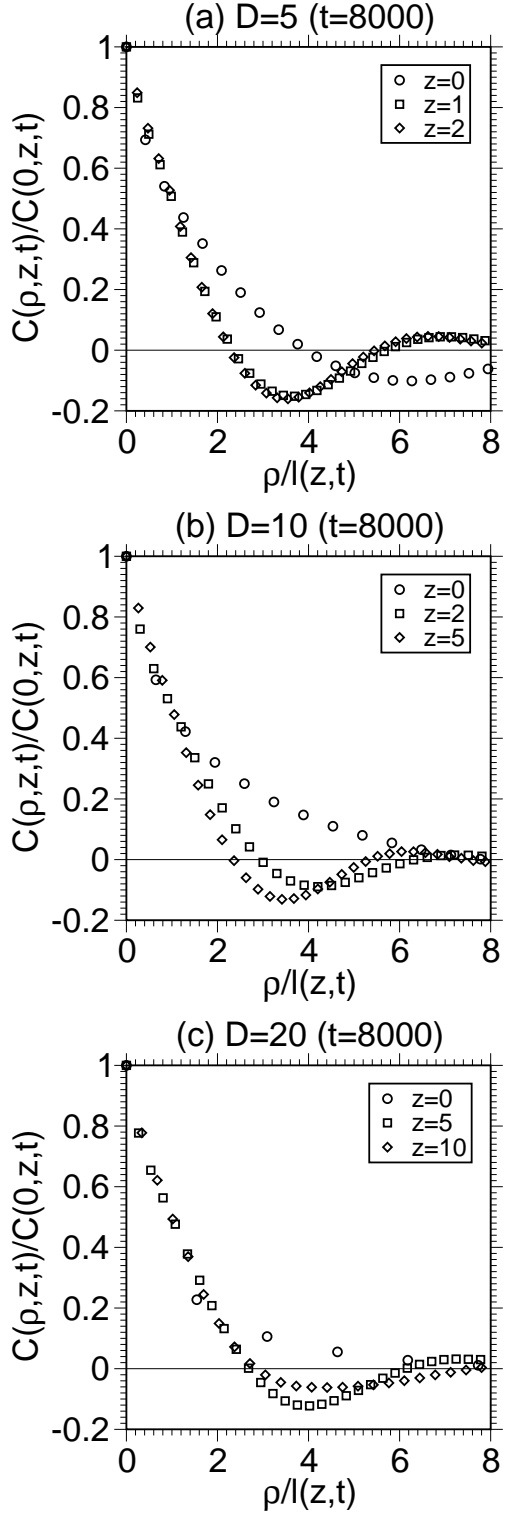
 
\centering
\includegraphics*[width=0.4\textwidth]{fig8a.eps} \\
\vskip0.2cm
\includegraphics*[width=0.4\textwidth]{fig8b.eps} \\
\vskip0.2cm
\includegraphics*[width=0.4\textwidth]{fig8c.eps}
\caption{\label{fig8}
Plot of normalized correlation function $C(\rho,z,t)/C(0,z,t)$ vs.
$\rho/\ell(z,t)$ at $t=8000$ for (a) $D=5$, (b) $D=10$, and (c) $D=20$.
We show data for several values of $z$ (distance from the left
wall), as indicated in the figure. The time $t=8000$ is
chosen such that it corresponds to the later stages of coarsening.}
\end{figure}

\begin{figure}
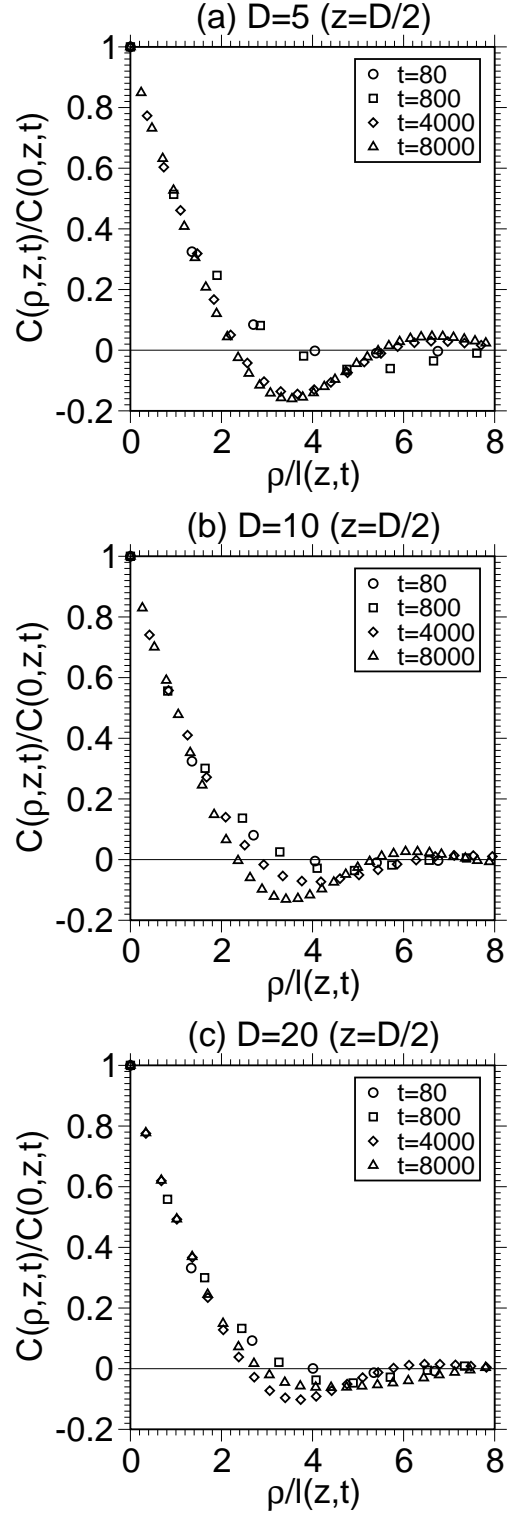
 
\centering
\includegraphics*[width=0.4\textwidth]{fig9a.eps} \\
\vskip0.2cm
\includegraphics*[width=0.4\textwidth]{fig9b.eps} \\
\vskip0.2cm
\includegraphics*[width=0.4\textwidth]{fig9c.eps}
\caption{\label{fig9}
Plot of normalized correlation function $C(\rho,z,t)/C(0,z,t)$
vs. $\rho/\ell(z,t)$ for $z=D/2$, and (a) $D=5$, (b) $D=10$,
and (c) $D=20$. We show data for times $t=80,800,4000,8000$
in all cases, as indicated by the different symbols.}
\end{figure}

We use this model to elucidate all the main characteristics of SDSD.
The surfaces become the origin of SDSD waves, which consist of alternating
enrichment and depletion layers of the preferred component A. These
waves coalesce in the central region of the film, giving rise to a
layered structure -- see the $t=800$ profile in Fig.~\ref{fig4}b and the
$t=4000$ profile in Fig.~\ref{fig4}d.  This layered state subsequently
breaks up into a columnar structure that coarsens laterally -- see
the cross-sections in Figs.~\ref{fig6} and \ref{fig7}. Therefore, the
local concentration of A-particles at the walls grows rapidly at first,
resulting in rather large values at early times, followed by a decrease
at later times (Fig.~\ref{fig5}).

\begin{figure}
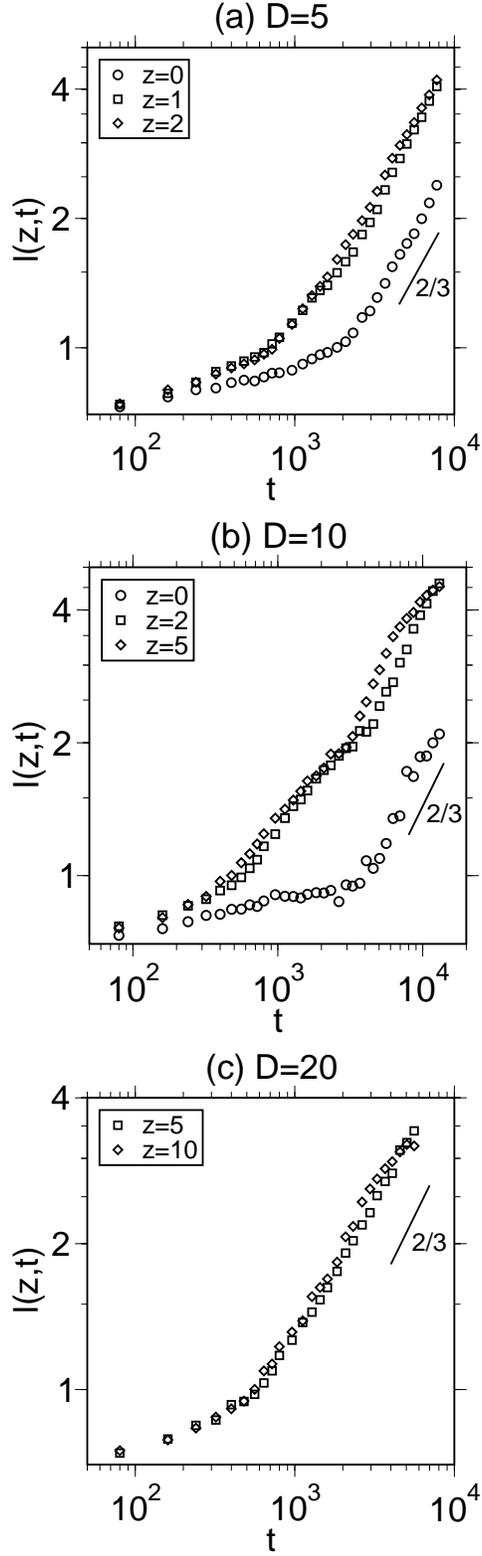
 
\centering
\includegraphics*[width=0.38\textwidth]{fig10a.eps} \\
\vskip0.2cm
\includegraphics*[width=0.38\textwidth]{fig10b.eps} \\
\vskip0.2cm
\includegraphics*[width=0.38\textwidth]{fig10c.eps}
\caption{\label{fig10}
Time-dependence of layer-wise length scale $\ell(z,t)$,
plotted on a log-log scale. We show data for different values
of $z$, and (a) $D=5$, (b) $D=10$, (c) $D=20$. The straight
lines have a slope of 2/3.}
\end{figure}

In the initial stages of phase separation, domain growth in the film
interior resembles spinodal decomposition in bulk mixtures, where a
bicontinuous percolating structure forms at compositions near the
critical concentration. During this stage, $\ell(z,t)$
grows only rather slowly (Fig.~\ref{fig10}). In this regime,
the concentration of A in the film center decreases, due to the
growth in thickness of the surface enrichment layers. Therefore, the
percolating structure breaks up into separated A-rich droplets which have
a cylindrical shape, with height of order $D$ in the $z$-direction and
radius of order $\ell(z,t)$. However, these droplets are connected
through the A-rich enrichment layers at the walls. The observation
that, in this droplet growth stage, the local concentration at the surface
decreases can be understood as follows. For the chosen parameters,
the B-rich phase exhibits incomplete wetting of A at the walls -- for
complete wetting, no overshoot of $\psi_{\rm av} (z=0,t)$ vs. $t$
would be expected in Fig.~\ref{fig5}. A remarkable feature of our
results is that domain growth is compatible with the inertial growth
law, $\ell(t) \propto t^{2/3}$, during the droplet growth stage
(Fig.~\ref{fig10}). This is reminiscent of studies \cite{50} of phase
separation in strictly 2-$d$ fluids, where Langevin equations including
hydrodynamic interactions were simulated.

\begin{figure}
\centering
\includegraphics*[width=0.9\textwidth]{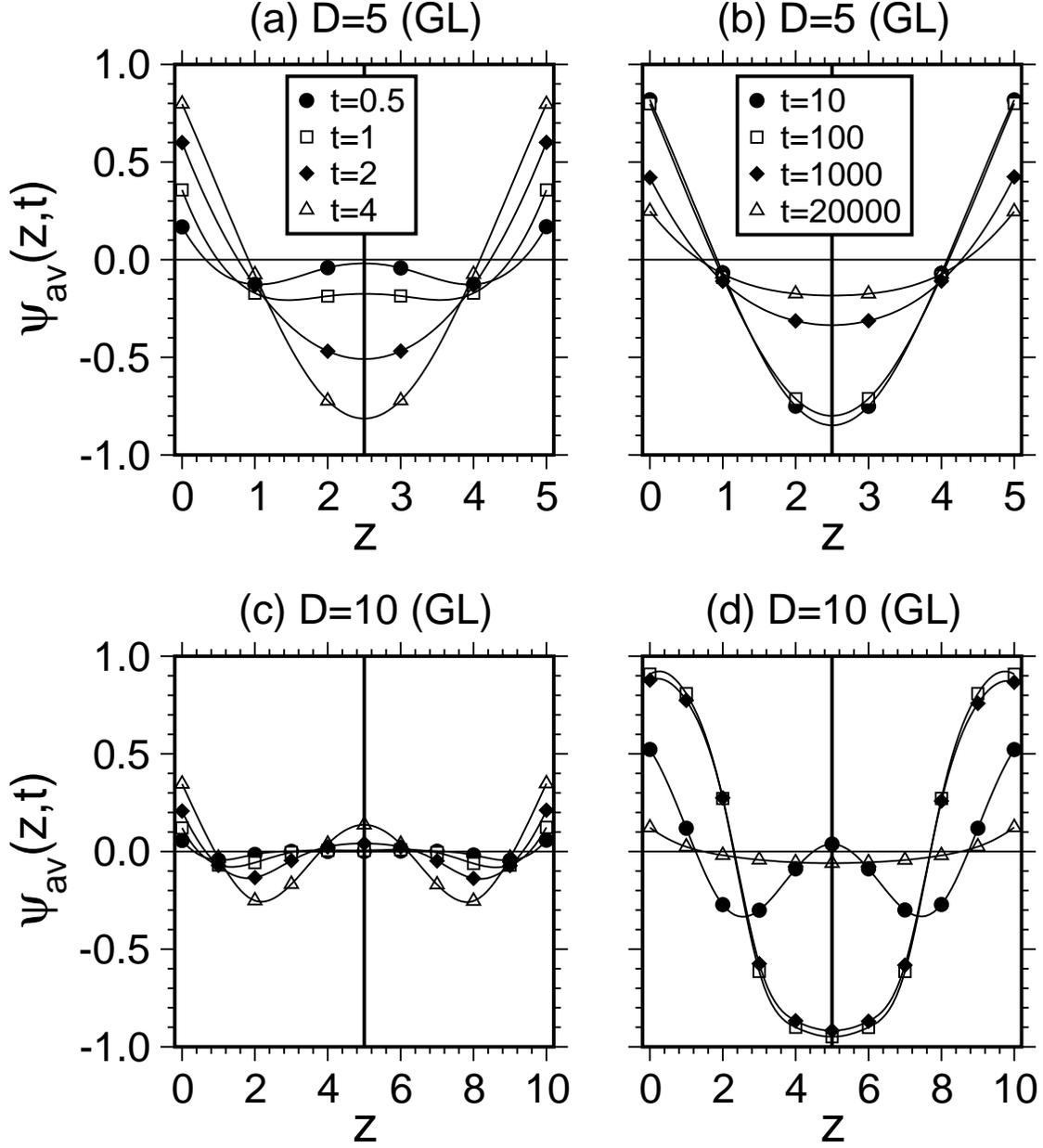}
\caption{\label{fig11}
Laterally averaged order parameter profiles, $\psi_{\rm av} (z,t)$
vs. $z$, obtained from the GL simulations described in Sec.~II.C \cite{40}.
The GL data was obtained as an average over 5 independent runs with $L=256$. 
We show data for films of thickness (a) $D=5$ at early times
($t=0.5,1,2,4$); (b) $D=5$ at late times ($t=10,100,1000,20000$);
(c) $D=10$ at early times ($t=0.5,1,2,4$); (d) $D=10$ at late times
($t=10,100,1000,20000$). The symbol usage is the same for (a),(c)
as well as for (b),(d).}
\end{figure}

It is clear that an extension of the brute force MD approach to much
larger linear dimensions (needed at later times) requires prohibitively
large amounts of computer time. On the other hand, our MD study does
reach mesoscopic length scales, which are significantly larger than
inter-particle distances. This suggests that our MD study should be
supplemented by Langevin studies of the Ginzburg-Landau (GL) models
described in Sec.~II.C, so as to enable a simulation extending from
microscopic to macroscopic scales. Let us briefly discuss a comparison
for the early stages of SDSD in thin films, where the hydrodynamic
interactions can be disregarded. Figure~\ref{fig11} is analogous to
Fig.~\ref{fig4}, but is taken from a GL simulation of model B with
appropriate boundary conditions \cite{40}. The details of this simulation
are discussed at the end of Sec.~II.C.  The qualitative similarity of
the evolution of the depth profiles in Figs.~\ref{fig4} and \ref{fig11}
is striking. For the GL simulations, the spatial degrees of freedom
were rather coarsely discretized (with $\Delta x=1$), and hence the
small-scale structure close to the walls cannot be resolved. Apart from
this difference, the GL profiles are in good agreement with those obtained
from the MD simulation for short times, if we equate the time units as
$t_{\rm GL}\simeq 160~t_{\rm MD}$.  Recall that the natural time-scale in
fluids is given in terms of the structural relaxation time \cite{103},
and the latter is of the same order as the shear viscosity, which is
$\eta \simeq 7$ LJ units in this case. Since the MD time unit $t_{\rm
MD}\equiv t_0 \simeq 1/\sqrt{48} \simeq 1/7$ LJ units, we conclude that
the natural fluid time-scale is about 50 MD time units.

The qualitative behavior of the GL results \cite{40} (formation of
surface enrichment layers and a metastable layered state $\rightarrow$
break up due to lateral phase separation $\rightarrow$ coarsening of
columnar structures) is similar to the results of the present MD study.
However, in quantitative respects, the intermediate stages of coarsening
are rather different for the GL model. When the layered state breaks
up into a laterally inhomogeneous state in the GL model, one often
encounters a period of time where $\ell$ stays roughly constant, or
even decreases. We do not observe such a transient behavior in the MD
runs -- as a matter of fact, the layered state does not survive for any
appreciable period of time. Further, asymptotic domain growth in the GL
model is compatible with an $\ell(t) \propto t^{1/3}$ law, as predicted
by the Lifshitz-Slyozov theory. On the other hand, our MD results are
consistent with a growth law $\ell(t) \propto t^{2/3}$. Clearly, a GL
study of SDSD in thin films, which incorporates hydrodynamic interactions,
would be very desirable. Further, a detailed comparison of the present MD
results with corresponding lattice Boltzmann studies should be worthwhile,
but is left to future studies. Finally, we hope that the present work
will further stimulate experimental studies of phase separation in thin
films. \\

\noindent Acknowledgments: We thank the Deutsche
Forschungsgemeinschaft (DFG) for support via grant No Bi 314/18-2
(SKD), the Emmy Noether Program (JH), and grant SFB 625/A3 (SP).

\newpage

\end{document}